# Design of a dual species atom interferometer for space

*Thilo Schuldt[1], Christian Schubert[2], Markus Krutzik[3], Lluis Gesa Bote[4], Naceur Gaaloul[2], Jonas Hartwig[2], Holger Ahlers[2], Waldemar Herr[2], Katerine Posso-Trujillo[2], Jan Rudolph[2], Stephan Seidel[2], Thijs Wendrich[2], Wolfgang Ertmer[2], Sven Herrmann[5], André Kubelka-Lange[5], Alexander Milke[5], Benny Rievers[5], Emanuele Rocco[6], Andrew Hinton[6], Kai Bongs[6], Markus Oswald[7], Matthias Franz[7], Matthias Hauth[3], Achim Peters[3], Ahmad Bawamia[8], Andreas Wicht[8], Baptiste Battelier[9], Andrea Bertoldi[9], Philippe Bouyer[9], Arnaud Landragin[10], Didier Massonnet[11], Thomas Lévèque[11], Andre Wenzlawski[12], Ortwin Hellmig[12], Patrick Windpassinger[12,1], Klaus Sengstock[12], Wolf von Klitzing[13], Chris Chaloner[14,2], David Summers[14], Philip Ireland[14], Ignacio Mateos[4], Carlos F. Sopuerta[4], Fiodor Sorrentino[15], Guglielmo M. Tino[15], Michael Williams[16], Christian Trenkel[16], Domenico Gerardi[17], Michael Chwalla[17], Johannes Burkhardt[17], Ulrich Johann[17], Astrid Heske[18], Eric Wille[18], Martin Gehler[18], Luigi Cacciapuoti[18], Norman Gürlebeck[5], Claus Braxmaier[1,5], Ernst Rasel[2]*

[1] German Aerospace Center (DLR), Institute of Space Systems, Robert-Hooke-Str. 7, 28359 Bremen, Germany
[2] Institut für Quantenoptik, Leibniz Universität Hannover, Welfengarten 1, 30167, Hannover, Germany
[3] Humboldt-Universität zu Berlin, Institut für Physik, Newtonstr. 15, 12489 Berlin, Germany
[4] Institut de Ciències de l'Espai (CSIC-IEEC), Campus UAB, Facultat de Ciències, 08193 Bellaterra, Spain
[5] Zentrum für angewandte Raumfahrttechnologie und Mikrogravitation (ZARM), Universität Bremen, Am Fallturm, 28359 Bremen, Germany
[6] School of Physics and Astronomy, University of Birmingham, Birmingham, B152TT, United Kingdom
[7] Institut für Optische Systeme, University of Applied Sciences Konstanz (HTWG), Brauneggerstr. 55, 78462 Konstanz, Germany
[8] Ferdinand-Braun-Institut, Gustav-Kirchhoff-Str. 4, 12489 Berlin, Germany
[9] Laboratoire Photonique, Numérique et Nanosciences-LP2N Université Bordeaux-IOGS-CNRS: UMR 5298, Talence, France
[10] LNE-SYRTE, Observatoire de Paris, CNRS and UPMC, 61 avenue de l'observatoire, 75014 PARIS, France
[11] CNES - Centre National d'Études Spatiales, 18 Avenue Édouard Belin, 31400 Toulouse, France
[12] Institut für Laserphysik, Universität Hamburg, Luruper Chaussee 149, 22761 Hamburg, Germany
[13] Institute of Electronic Structure and Laser, Foundation for Research and Technology - Hellas, Vassilika Vouton, GR-71110 Heraklion, Greece
[14] SEA House, Bristol Business Park, Coldharbour Lane, Bristol BS16 1EJ, United Kingdom
[15] Dipartimento di Fisica e Astronomia and LENS, Università di Firenze - INFN, Sezione di Firenze - via G. Sansone 1, 50019 Sesto Fiorentino (Firenze), Italy
[16] Astrium Ltd, Gunnels Wood Road, Stevenage SGI 2AS, United Kingdom
[17] Astrium GmbH - Satellites, Claude-Dornier-Str., 88090 Immenstaad, Germany
[18] ESA - European Space Agency, ESTEC, Keplerlaan 1, 2200 AG Noordwijk ZH, Netherlands

*Corresponding author:* Thilo Schuldt, tel. +49 7531/206-379, fax +49 7531/206-521, thilo.schuldt@dlr.de

*Abstract* Atom interferometers have a multitude of proposed applications in space including precise measurements of the Earth's gravitational field, in navigation & ranging, and in fundamental physics such as tests of the weak equivalence principle (WEP) and gravitational wave detection. While atom interferometers are realized routinely in ground-based laboratories, current efforts aim at the development of a space compatible design optimized with respect to dimensions, weight, power consumption, mechanical robustness and radiation hardness. In this paper, we present a design of a high-sensitivity differential dual species $^{85}$Rb/$^{87}$Rb atom interferometer for space, including physics package, laser system, electronics and software. The physics package comprises the atom source consisting of dispensers and a 2D magneto-optical trap (MOT), the science chamber with a 3D-MOT, a magnetic trap based on an atom chip and an optical dipole trap (ODT) used for Bose-Einstein condensate (BEC) creation and interferometry, the detection unit, the vacuum system for $10^{-11}$ mbar ultra-high vacuum generation, and the high-suppression factor magnetic shielding as well as the thermal control system. The laser system is based on a hybrid approach using fiber-based telecom components and high-power laser diode technology and includes all laser sources for 2D-MOT, 3D-MOT, ODT, interferometry and detection. Manipulation and switching of the laser beams is carried out on an optical bench using Zerodur bonding technology. The instrument consists of 9 units with an overall mass of 221 kg, an average power consumption of 608 W (819 W peak), and a volume of 470 liters which would well fit on a satellite to be launched with a Soyuz rocket, as system studies have shown.

---

[1] Now at Johannes-Gutenberg-University Mainz, Staudingerweg 7, 55099 Mainz, Germany
[2] Now at Trym Systems Ltd. 1 College Park Drive Westbury-on-Trym, Bristol BS10 7AN, United Kingdom

# 1. Introduction

In analogy to optical interferometry, atomic matter waves can be split and recombined, resulting in an interference signal, which can be utilized in various applications. Atom interferometers (AI) can be used as high-precision sensors for acceleration and rotation [1][2][3][4], enabling six-axes inertial sensing [5] and e.g. precision measurements of the gravitational acceleration [6][7] and the Earth's gravity gradient [8]. Furthermore, atom interferometers enable precision tests of fundamental physics: The weak equivalence principle (WEP) can be tested by measuring the differential acceleration of two different atom species using a differential AI with high common mode suppression [9]. Proposed gravitational wave detectors use ballistic atoms in AI setups as inertial test masses [10][11].

All application areas detailed above favor AI operation in space. Compared to laboratory, drop tower or sounding rocket experiments, space offers a zero-g environment, orders of magnitude longer integration times, larger variations of gravitational potential and velocities, orders of magnitude better relative uncertainties and a well-controlled micro-vibration environment. A space-based Earth gravity gradiometer using macroscopic test masses for gravity gradient determination was operated aboard the GOCE satellite [12] launched in 2009, following up the gravity measurement missions CHAMP [13] and GRACE [14]. Quantum gravity gradiometers using atoms as inertial test masses, where the effect caused by Earth's gravity is measured using atom interferometric techniques, were proposed [15][16]. Such devices are expected to offer a higher accuracy than GOCE [16].

A space-based test of the weak equivalence principle (WEP) using macroscopic test masses is currently in the implementation phase, cf. the ESA-CNES collaborative mission Microscope (MICROSatellite à Trainée Compensée pour l'Observation du Principe d'Équivalence) [17]. WEP tests using quantum matter are proposed and currently under assessment at ESA. STE-QUEST (Space-Time Explorer and Quantum Equivalence Principle Space Test) was an M3 mission candidate within ESA's cosmic vision program with launch opportunity between 2022 and 2024 utilizing a dual species $^{85}$Rb/$^{87}$Rb differential atom interferometer flown in a highly elliptical Earth orbit [18][19][20]. Operating a similar AI instrument on the International Space Station (ISS) is proposed within the QWEP (Quantum Weak Equivalence Principle) project [21].

Most gravitational wave detectors operated on ground (such as GEO600, LIGO, VIRGO, TAMA) [22] and proposed for space (eLISA, evolved Laser Interferometer Space Antenna, DECIGO) [23] rely on long-baseline laser interferometry measuring changes in distance between distant macroscopic test masses. An alternative approach using atom interferometry was proposed and detailed during the last years: Two atom interferometers are operated on two distant spacecraft using common laser beams. In this scenario, the signal of a gravitational wave will be encoded in the differential phase shift [10][11].

While each AI space application has its specific requirements on the instrument design, the underlying technologies are similar. This includes, e.g. cold atom preparation, vacuum and laser technology as well as specific assembly-integration technologies for space compatibility. In this paper, we present the design of the dual species $^{85}$Rb/$^{87}$Rb differential atom interferometer as it was developed by a European consortium in the context of the STE-QUEST Phase A study. STE-QUEST proposes to test the WEP aboard a satellite by comparing the propagation of two Bose-Einstein condensates of $^{85}$Rb and $^{87}$Rb in the Earth's gravitational field using atom interferometer techniques. The AI design complies with the primary science objective of measuring the Eötvös ratio $\eta = |\Delta a/g|$ with an accuracy of 2 parts in $10^{15}$, where $\Delta a$ denotes the differential acceleration between the two isotopes and g the projection of the gravitational acceleration parallel to $\Delta a$. A comprehensive description of the STE-QUEST mission and its science objectives is given in [18][19][24].

To identify any schedule and development risks in a space mission, the technolgy maturity of all elements of such a mission needs to be assessed. This maturity is quantified as a so-called TRL – the technology readiness level. The highest level is 9, which corresponds to flight-proven elements. The lowest level is 1, where basic principles are observed and potential applications are identified, however, element concepts are not yet formulated. Levels 3 and 4 correspond to proof-of-concept and functional validation in a laboratory environment. At the beginning of the implementation phase of a space project a TRL of 5 or 6 is typically required. Critical functions of the elements have been verified on breadboards (5) or full models (6) for the relevant enviroment.

The paper is organized as follows: Chapter 2 first gives an overview over the operation principle of an atom interferometer while Chapter 3 comprises an overview over the overall instrument design, its budgets, and the corresponding key design criteria. The on-ground and in-space calibration procedures are highlighted in Chapter 4. The designs of the different units of the atom interferometer are then given in detail: physics package (Chapter 5), laser system (Chapter 6), and electronics & software (Chapter 7). Emphasis is put on specific issues relevant for space operation. This includes the use of specific space-qualified technologies, and in case space-qualified components, and design optimization with respect to mass, dimension, power consumption and interfaces to the spacecraft.

## 2. Operation Principle of an Atom Interferometer

Atom interferometry relies on the superposition principle and the wave particle duality, allowing to observe interference patterns of matter waves. Interferometer geometries analogous to the Mach-Zehnder interferometer in optics can be implemented by the coherent manipulation of neutral atoms via light fields as used in gravimeters based on atom interferometry [6][7]. The light fields are pulsed and act as beam splitters which coherently split, re-direct, and recombine the matter waves (Figure 1). During each atom light interaction the position of the atoms is referenced with respect to the wave fronts of the light fields. In a retro reflection set up the wave fronts are defined by a reference mirror. Effectively, the acceleration of the atoms with respect to the mirror is tracked. After the pulse sequence a phase shift of $\mathbf{k} \cdot \mathbf{a} T^2$ is imprinted onto the atoms where $\mathbf{k}$ denotes the effective wave vector, $\mathbf{a}$ the relative acceleration between atoms and mirror, and T the free evolution time between two subsequent beam splitter pulses. The phase shift is encoded in the transition probability of the two output ports of the interferometer. A measurement cycle consists of three steps: preparation of the atomic ensembles, coherent manipulation, and detection of the transition probabilities.

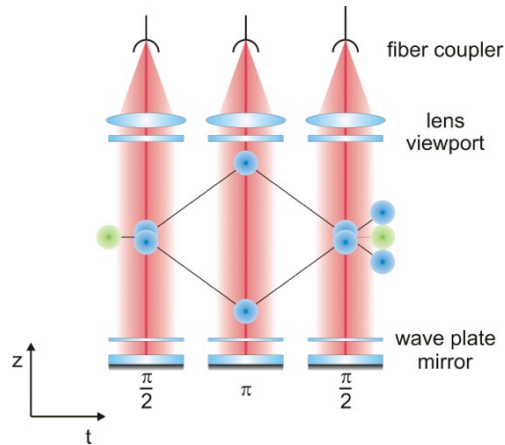

**Figure 1:** Beam splitter setup.

If two species are simultaneously interrogated, the differential acceleration Δa between the two can be measured. Subsequently, the Eötvös ratio η = |Δa/g| with the gravitational acceleration g can be determined. Herein, the shot noise limited sensitivity $s_{\Delta a}$ to Δa is given by the formula $s_{\Delta a}=(2/N)^{1/2}(CkT^2)^{-1}(T_c/t)^{1/2}$. It scales favorably with a high atom number N per species, a high interferometer contrast C, a short cycle time $T_c$, and especially a long free evolution time T. The averaging time is denoted by t. Space borne operation would enable long free evolution times T not accessible on ground.

The described set up in this paper is adapted to the species $^{87}$Rb and $^{85}$Rb like the WEP test experiments in [9][25]. This choice promises a high suppression ratio against acceleration noise and systematics [24] and inherits a comparably high maturity from zero-g [29][32][33] and ground based experiments [9][25][39]. Preparing the atoms as Bose-Einstein condensates (BECs) with low expansion rates ensures a high contrast and reduces systematic effects related to imperfect wave fronts. By combining an atom chip and an optical dipole trap (ODT) a fast BEC production is expected which allows for a short cycle time. The beam splitting is foreseen in a double diffraction scheme which inherently suppresses certain systematics as light shift and related noise figures [40]. A fluorescence detection via a CCD camera which spatially resolves both output ports of the interferometer simultaneously is chosen to suppress noise figure due detection laser frequency and intensity stability [43][26].

## 3. Instrument Overall Design

The AI instrument consists of three main functional units, subdivided into single physical boxes for spacecraft integration (cf. the functional diagram shown in Figure 2):

- Physics Package: one cylinder
- Laser System: 3 boxes
- Electronics: 5 boxes

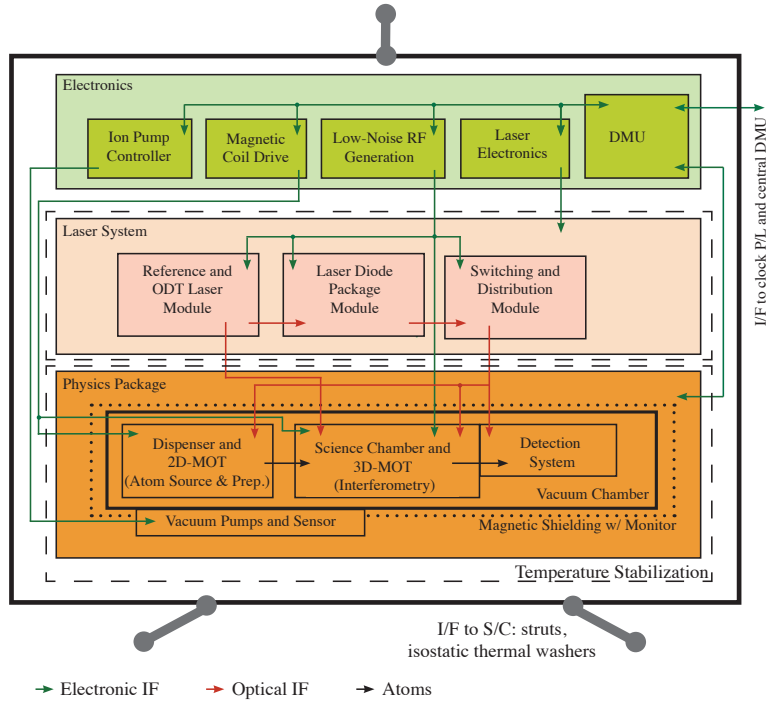

**Figure 2:** Functional diagram of the AI payload [18].

The *physics package* provides the environment for performing the interferometry measurements with the two species matter waves. This system comprises the coherent matter wave source, an atom chip combined with a dipole trap, the beam splitter unit comprising a highly stable, optically super flat retro reflector and the detection. The matter waves have to be generated and tracked in ultra-high vacuum and in a magnetically shielded environment with a precise and highly stable magnetic field.

The *laser system* comprises the laser sources for laser cooling and trapping, state preparation and detection as well as the coherent manipulation of the atoms for forming the interferometer. In addition, the system includes the laser for confining the atoms in a dipole trap and the light distribution module for modifying and controlling the laser frequency, polarization and power.

Both, the physics package and the laser system, require a sophisticated control *electronics system*. It consists of a data management unit (DMU), a low noise RF generator, an ion pump controller, and several driver modules to control the laser system and the physics package.

Each of these subsystems is separately temperature stabilized and linked by electrical and/or optical interfaces with each other. The AI instrument DMU is linked to the central spacecraft DMU via electronic (data and power) interfaces and the AI system is connected to the spacecraft via mechanical interfaces.

The AI design detailed in the following is based on strong European developments in this field, including the pre-phase A studies in the ELIPS-3 program Space Atom Interferometer (SAI) [27][28], Quantum gases in microgravity (SpaceBEC), the DLR project QUANTUS (QUANTengase Unter Schwerelosigkeit) [29][30] and the CNES project I.C.E. (Interférométrie Cohérente pour l'Éspace) [31][32]. The QUANTUS and I.C.E. projects both include dual species atom interferometer experiments, in case of QUANTUS carried out in a drop-tower (and on a sounding rocket by the end of 2014: MAIUS project, MAteriewellen Interferometer Unter Schwerelosigkeit), and in case of I.C.E. onboard a parabolic zero-g flight.

Taking into account the heritage of the aformentioned state-of-the-art experiments and specific requirements for operation in space, the AI design is further developed with respect to dimensions, mass and power consumption as well as mechanical robustness and radiation hardness. Specific issues on thermal management of the subsystems (especially the physics package) and interfaces to the spacecraft were investigated. Technology Readiness Levels (TRL) of single components were evaluated and trade-offs between different technologies were carried out. If possible, the AI design is planned with space-qualified components.

A budget overview of the AI subsystems is given in Table 1. The values are generated using a bottom-up approach starting from component level and taking into account the instrument CAD model. The total instrument mass is 221 kg, the total average power consumption 608 W and the peak power consumption 819 W.

**Table 1:** Budget overview over the AI subsystems. The values for mass and power all include a 20% component level margin, but no system level margin.

|  |  |  | DIMENSIONS | | MASS | POWER | |
|---|---|---|---|---|---|---|---|
|  |  |  |  |  |  | Average | Peak |
|  |  |  | Length x width x height or length x diameter (mm x mm x mm) | Volume (l) | incl. 20% margin (kg) | incl. 20% margin (W) | incl. 20% margin (W) |
| **Physics Package (Unit)** | PPU | Cylinder | 1000 x 660 | 342,2 | 134,6 | 73,8 | 156,8 |
| **Laser System** |  |  |  |  |  |  |  |
| Reference and ODT Laser | ROL | Box | 310 x 310 x 100 | 9,6 | 8,5 | 42,6 | 62,6 |
| Diode Laser Package Unit | DLP | Box | 400 x 390 x 200 | 31,2 | 26,4 | 53,3 | 53,3 |
| Switching and Distribution Module | SDM | Box | 400 x 370 x 125 | 18,5 | 16,8 | 7,2 | 20,6 |
| **Electronics** |  |  |  |  |  |  |  |
| Data Management Unit | DMU | Box | 300 x 300 x 300 | 27,0 | 12,7 | 153,8 | 155,2 |
| Magnetic Coil Drive Electronics | MDE | Box | 300 x 300 x 200 | 18,0 | 7,1 | 95,8 | 211,6 |
| RF Generation | RFG | Box | 300 x 300 x 150 | 13,5 | 7,2 | 88,8 | 88,8 |
| Laser Control Electronics | LCE | Box | 300 x 250 x 100 | 7,5 | 6,1 | 91,2 | 91,2 |
| Ion Pump Controller | IPC | Box | 200 x 100 x 100 | 2,0 | 1,2 | 1,8 | 1,8 |
| **Subtotal** |  |  |  | **469,5** | **220,7** | **608,1** | **814,3**[1] |

The key parameters underlying the AI design are given in Table 2. It represents a dual species atom interferometer based on the two isotopes of rubidium, $^{85}$Rb and $^{87}$Rb, measuring the differential acceleration of the two atomic isotopes along one axis for performing a WEP test. For achieving an optimum suppression of common-mode noise, which is related to basically all possible disturbance sources, the two atomic species are simultaneously prepared, coherently manipulated and detected with optimally overlapped atomic clouds of the two species. The design facilitates a number of $10^6$ atoms and a free evolution time between the beamsplitter pulses of T = 5s. With these parameters the integrated sensitivity to the Eötvös parameter will be $5.2 \cdot 10^{-14}$ per orbit, and the target sensitivity at the $2 \cdot 10^{-15}$ level will be reached after integrating over 1.5 years [18][24]. The experimental sequence including BEC gene ration and atom preparation is shown in Figure 3 and detailed in [18]. The preparation phase is followed by the interferometer pulse sequence and the detection sequence.

**Table 2:** AI key specifications. The interferometer contrast is depicted by $C \geq 0.6$, which changes dependent on the satellite attitude. The integration time is given by $\tau$. The Eötvös ratio $\eta = |\Delta a/g|$ is determined by dividing the differential acceleration signal $\Delta a$ by the projection of the local gravitational acceleration g onto the sensitive axis with values between 3 – 8 m/s$^2$.

| Atomic species | $^{85}$Rb / $^{87}$Rb |
|---|---|
| Sensitivity to accelerations |  |
| - single shot / single interferometer | $1.24 \cdot 10^{-12}$ m/s$^2$ $\cdot$C$^{-1}$ $\leq 2.07 \cdot 10^{-12}$ m/s$^2$ |
| - single shot / differential | $1.75 \cdot 10^{-12}$ m/s$^2$ $\cdot$C$^{-1}$ $\leq 2.92 \cdot 10^{-12}$ m/s$^2$ |
| - differential integration behavior | $7.83 \cdot 10^{-12}$ m/(s$^2$Hz$^{1/2}$) $\cdot$C$^{-1}$ $\cdot \tau^{-1/2}$ |
|  | $\leq 13.05 \cdot 10^{-12}$ m/(s$^2$Hz$^{1/2}$) $\cdot \tau^{-1/2}$ |
| Atom number (each species) | $10^6$ |
| Free evolution time (2T) | 10 s |
| Experiment cycle time | 20 s |
| Magnetic shielding factor | > 10.000 |

---

[1] The total peak power is calculated for the time interval with the highest power consumption of the whole AI instrument, in this case for the interval between 2 s and 2.1 s of the 20 s experimental cycle (corresponding to molasses cooling and chip trap loading, cf. Figure 2).

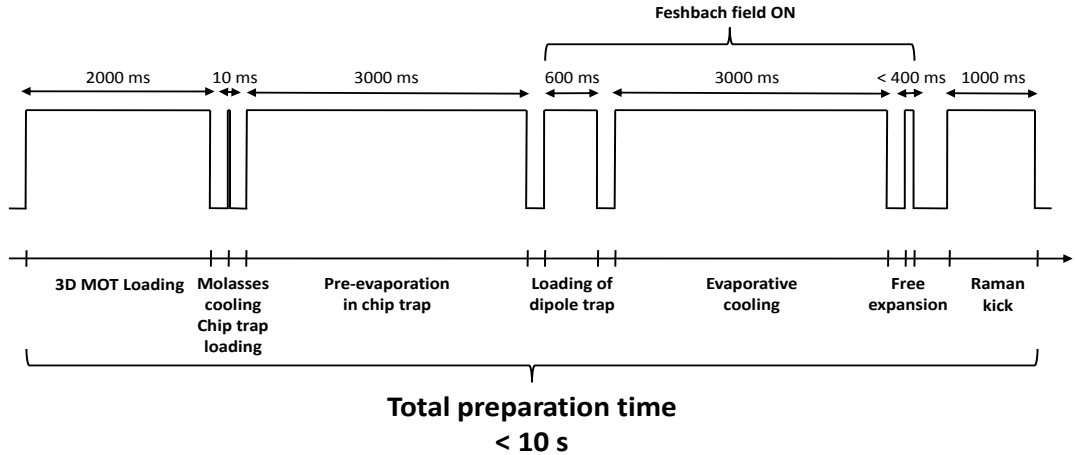

**Figure 3:** Experiment cycle sequence for BEC generation and atom preparation [18]. The preparation phase lasts approx. 10 s, followed by an interferometer pulse sequence and a detection sequence lasting another 10 s. The Raman kick sequence is applied to move the atomic ensembles 10.5 mm away from the atom chip. It consists of two subsequent Raman beam splitter pulses applied for each isotope.

## 4. Calibration

Nominal performance is only possible in space borne operation. Consequently, the procedures on ground will be limited to calibrate the response of photodiodes, magnetometers, thermistors, the output of control electronics, and the CCD cameras for detection system, also with respect to interspecies cross talk. The source operation will be initialized with reduced performance on ground. Subsequently, these parameters will be the preset for further optimization in space. During the first 4 months of the mission the source will be autonomously optimized with a differential evolution algorithm, beam splitter parameters will be adjusted, and systematics will be assessed. After entering the phase of nominal operation, science data will be taken during the perigee pass (0.5 h), a null measurement at apogee (0.5 h) will be used to remove systematics stable during one revolution. Remaining time per orbit will be used to gather additional data for further assessing systematics. Specifically the overlap of the two ensembles will be evaluated by a spatially resolved detection. This interleaved operation will enable a verification of the results throughout the whole mission.

## 5. Physics Package Design

Inside the physics package, $^{87}$Rb and $^{85}$Rb atoms are manipulated by laser light and magnetic fields to form the atom interferometer. The design of the setup is largely based on drop tower experiments with $^{87}$Rb BEC interferometers [29][30][33] and is adapted for a space mission with an interferometer baseline of 12 cm and simultaneous operation with $^{87}$Rb and $^{85}$Rb. The operation of the atom interferometer requires a vacuum system which is surrounded by telescopes for the application of laser light and several coils for providing offset fields. Three CCD cameras close to the main chamber are used for detection and calibration purposes (see Figure 6). Apart from the vacuum pumps, the whole physics package is shielded against magnetic stray fields by a four layer μ-metal shield. An overview sketch is shown in Figure 4. All parts of the physics package are mounted directly or indirectly to the baseplate which serves as mechanical and thermal connection to the spacecraft. The vacuum pumps are directly mounted while the science chamber and the μ-metal shield are connected via struts.

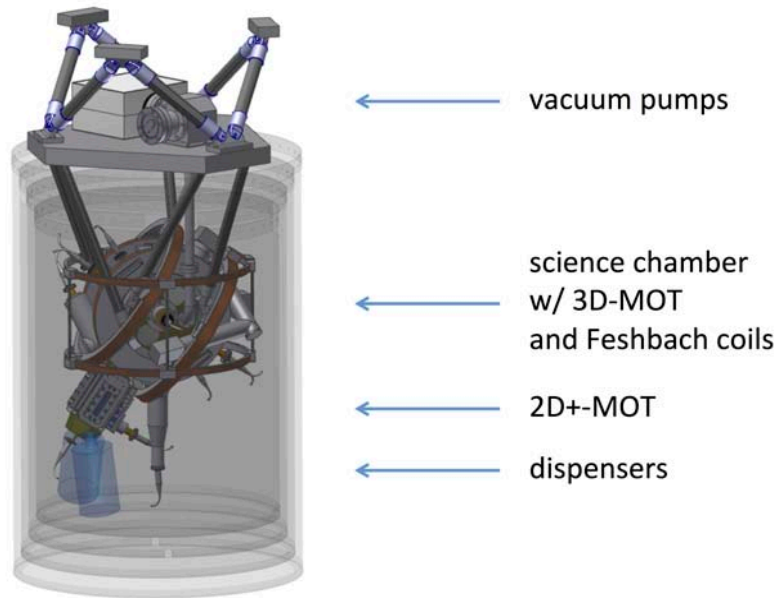

**Figure 4:** Overview drawing of the physics package.

## 5.1 Atom Source

The atomic source delivers pre-cooled atoms to the science chamber for further manipulation with a required flux of $10^{10}$ $^{87}$Rb atoms/s and $10^{9}$ $^{85}$Rb atoms/s. Its two main parts are a 2D+MOT (magneto-optical trap, see [34]) and two redundant dispensers as depicted in Figure 5. The dispensers are based on the already space-qualified Cs source design for the cold atom PHARAO space clock [35][36]. Inside a cylindrical tank, a porous titanium matrix stores the atoms. A mechanical valve chamber and a heater control the atomic flux. This will be adapted from Cs to provide $^{87}$Rb / $^{85}$Rb background vapor inside the 2D+MOT chamber. Each dispenser will deliver both Rb isotopes, two dispensers are foreseen for redundancy. No major issues concerning e.g. size and matrix material for a redesign to $^{87}$Rb / $^{85}$Rb are expected.

Subsequently, the 2D+MOT (Figure 5) forms a continuous beam of cold $^{87}$Rb and $^{85}$Rb atoms simultaneously which is fed through a differential pumping stage into the science chamber. Two counter propagating laser beam pairs combined with the magnetic fields produced by two anti-Helmholtz coil pairs cool and trap the atoms in two axes. The resulting atomic flux in the third axis depends on the trapping volume and the optical power. Sufficient performance is expected from the proposed design with two quadruples of 18 mm beams next to each other. Two optical fibers deliver cooling light to the 2D+MOT. Behind the fiber plug, the beams are widened and collimated with telescopes and then split using two prisms and a mirror to extend the trapping volume (Figure 5, middle/right). After passing through the chamber, they are reflected by a mirror with a λ/4 retardation coating, realizing two counter propagating laser light fields. Additional pushing and retarding beams on the axis of the atoms are supplied via dedicated telescopes to increase the flux. The coils have a rectangular shape and are located around the viewports.

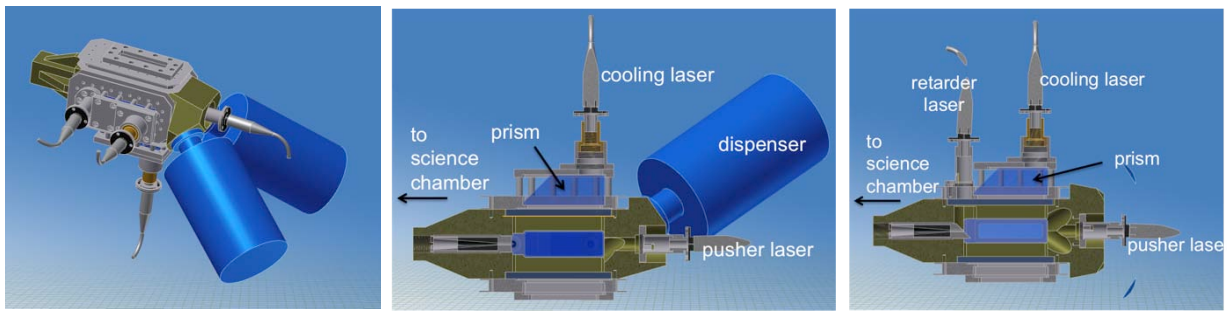

**Figure 5:** Overview drawing of the 2D+MOT including dispenser (left), cut through the 2D+MOT (middle, right). The coils are integrated inside the rectangular mounting structures (gray) for the telescopes and mirrors. Both isotopes are manipulated with the same coil pairs.

Comparable 2D+MOT set-ups are operated in the lab with $^{87}$Rb with magnetic field gradients of 19 G/cm. The flux is sufficient to reach loading rates of the 3D-MOT of $1.4\cdot10^9$ atoms per second [37]. By increasing the trapping volume and the provided laser power a higher rate is expected in the proposed set up. A version with smaller view ports for the cooling beams was vibration tested with up to 5.4 g$_{rms}$. Therein, all components are suited for operation with both $^{87}$Rb and $^{85}$Rb.

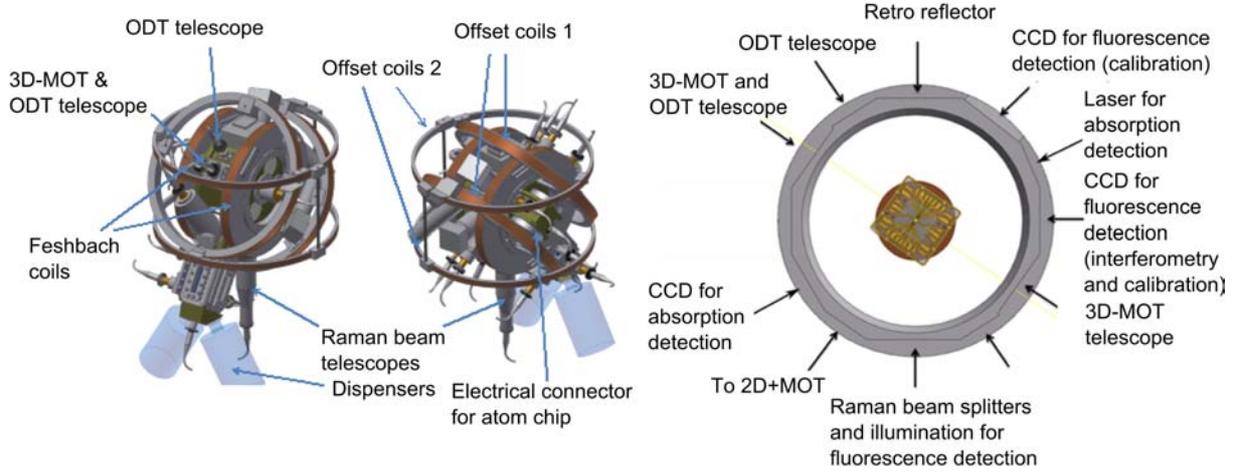

**Figure 6:** Overview drawing of the science chamber (left, middle) and viewport assignment (right). One CCD camera shall measure absorption images during the source calibration. The optical axis has the same distance of 4.5 mm to the surface of the atom chip as the ODT which enables in situ and time of flight measurements without further manipulation steps as the Raman kick. Two more CCD cameras will take fluorescence images from two different directions to obtain 3D spatial information about the overlap. These two optical axes enclose an angle of 60° and have distance of 15 mm to the atom chip in the plane of the sensitive axis of the atom interferometer. Opposing to the atom chip, the Raman kick telescope is mounted.

## 5.2 Science Chamber

The science chamber has to support a rapid dual species BEC production in 9 s with $10^6$ atoms per species and effective temperatures of 70 pK, interferometry with 2T = 10 s implying a baseline of 12 cm, and a shot noise limited detection system. To avoid inaccuracies in the measurements, magnetic field gradients have to be suppressed to below 500 μG/m, and the retro reflection mirror for the beam splitters has to comply with a peak to valley surface quality of λ/50.

In and around the science chamber, all elements for simultaneously generating $^{87}$Rb / $^{85}$Rb BECs and their coherent manipulation are assembled. The chamber features an atom chip / ODT hybrid trap for swift BEC generation [38] and optical access for coherent manipulation and detection. Loaded from the 2D+MOT [34], the atom chip generates magnetic fields for a 3D-MOT and ensures high transfer efficiency into the ODT via pre-evaporation in a magnetic trap. Evaporation to the BEC follows inside the ODT where a Feshbach field prevents a collapse of the $^{85}$Rb BEC [39].

The proposed science chamber will feature a dodecagon design as depicted in Figure 6 (right). To accommodate the interferometer with a baseline of 12 cm the inner diameter of the chamber will be 15.5 cm. At the centre of Figure 6 (right) the atom chip is shown. The orientation of the atom chip defines the light field axes for the 3D-MOT and ODT. The 3D-MOT consists of three pairs of laser beams perpendicular to each other, where each pair of laser beams consists of two counter-propagating laser beams. In the proposed setup, a 3D-MOT is implemented with the atom chip acting as mirror, to generate two of the counter-propagating laser beams [28]. The CAD model is shown in Figure 7. Two beams are sent to the atom chip under an angle of 45°; the third pair of laser beams is sent through the corresponding viewports of the science chamber parallel to the surface of the atom chip (compare Figure 6). The laser beams have a $1/e^2$ diameter of 20 mm.

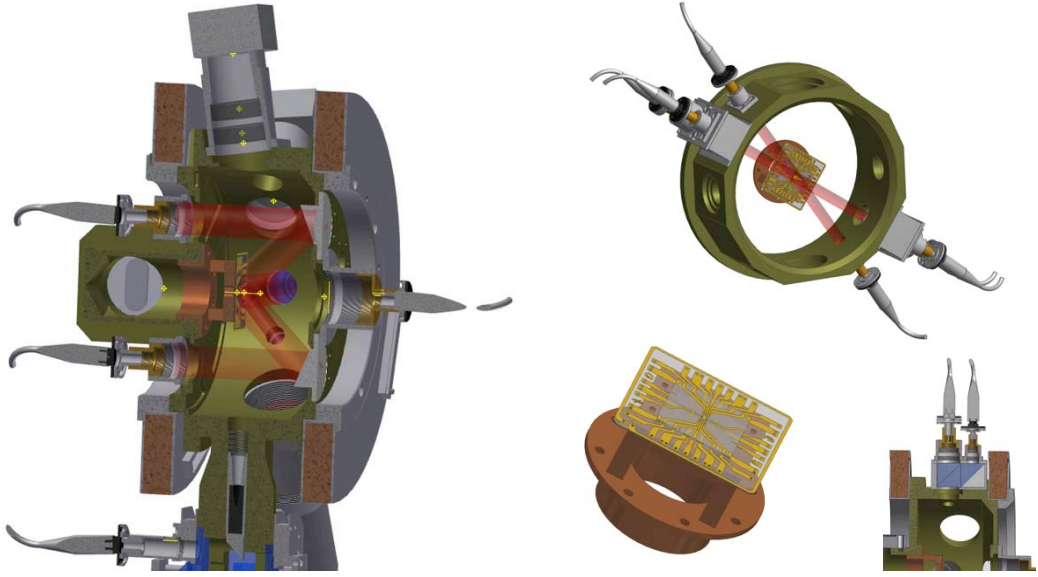

**Figure 7:** Atom chip MOT beam setup (left), atom chip (middle), ODT beam setup (top), and telescope setup for ODT / 3D-MOT beams (right).

The CAD model of the proposed atom chip is shown in Figure 7. The atom chip consists of three layers: The top two layers, called base and science layer are each formed by an array of micro fabricated gold wires on an AlN-substrate of 0,375 inch thickness. The bottom layer, called mesoscopic layer, is formed by Kapton clad copper wires. The size of the structures available with this approach varies from a few millimeters to a few micrometers. Therefore both large and strongly confined traps with high trap frequencies can be created. The major advantage of the proposed design is the low power consumption compared to macroscopic traps created purely by coils, while still having a large initial trap size and therefore high atom numbers. Furthermore, the high trap frequencies realizable with the design allow for a swift evaporation.

The surface of the top layer of the chip is coated with a highly reflective, dielectric layer. This reflective layer is used during the three dimensional magneto-optical trap: By using two reflected laser beams at 45° with respect to the chip surface and two laser beams in the plane of the chip surface in combination with a quadruple magnetic field from the chip and an offset field from the offset coil 1, the atoms coming from the 2D-MOT are trapped and cooled in all three directions. The specifications of the structures on the atom chip can be found in Table 3.

**Table 3:** Specifications of the atom chip.

| Layer | Structure sizes | Wire size | Rated Current |
|---|---|---|---|
| Mesoscopic U | 30 mm | 1 mm$^2$ | 7.5 A |
| Mesoscopic H | 12 mm | 1 mm$^2$ | 10 A |
| Base-Chip | 2.5 - 7.5 mm | 500 μm · 10 μm | 5 A |
| Science-Chip | 0.55 - 2.2 mm | 50 μm · 10 μm | 2 A |

Additionally, RF frequencies for evaporation in a range of 0-40 MHz and micro wave pulses tunable by ±10 MHz around 3.036 GHz and 6.835 GHz for pulse durations of 0-5 ms will be generated by the chip structures.

**Table 4:** Specifications of the magnetic coils. The Feshbach and offset 2 coils are both in Helmholtz configuration; the 2D-MOT coils have a rectangular shape and anti-Helmholtz configuration. In case of the offset 1 coil pair, the distance between the coils of 125 mm slightly differs from the ideal Helmholtz configuration.

| Coil | Radius | Windings | Magnetic Field |
|---|---|---|---|
| Feshbach | 103 mm | 400 | 34.92 G /A |
| Offset 1 | 150 mm | 108 | 7.12 G/A |
| Offset 2 | 170 mm | 20 | 1.06 G/A |
| 2D-MOT | 46 mm · 60 mm | 108 | 11.78 G/(cm·A) |

The homogenous magnetic fields used for the experiment are created using three pairs of coils in Helmholtz configuration (see Figure 6). They are made out of wirings of Kapton isolated copper wires on a plastic holder, for detailed specifications see Table 4. During the interferometry, one coil pair (Offset 2) generates a magnetic field of 1 mG to define the axis of the spin polarization. All other coils and chip structures are switched off at this point. Residual magnetic field gradients have to be below 500 µG/m over the baseline of the atom interferometer (see Chapter 5.5). A more stringend requirement exists after releasing the condensates from the trap when the atoms are still in magnetic sensitive states and the Feshbach field is still on. Until the switch off and the transfer to magnetic insensitive states, magnetic field gradients have to be below 3 µG/m (see Chapter 5.6).

Two laser beams enclosing an angle of 22.5° realize the optical dipole trap, see Figure 6, right, and Figure 7, top. One ODT laser beam uses the same viewport as the 3D-MOT laser beam; both beams are superimposed using a prism as detailed in Figure 7, right. Both ODT beams are focused to a have a minimum waist of 100 µm in a distance of 4.5 mm with respect to the center of the chip surface.

In the QUANTUS II project, a similar science chamber housing an atom chip and using the same telescope design is currently operated in the lab. It is specifically adapted for tests in the drop tower. Moreover, a second, comparable science chamber was manufactured for the sounding rocket mission MAIUS and withstood vibration tests at 5.4 $g_{rms}$. Both of these experiments feature an atom chip. The atom chip in the proposed design requires a slight modification since a micro wave antenna shall be included. The possibility to include such an antenna on a chip was demonstrated in other atom chip experiments.

Beam splitting is performed in a double diffraction scheme (cf. Figure 1) [40]. This kind of interferometers is intrinsically insensitive to the first order AC Stark shift, the Zeeman effect if no gradients are present, and Raman laser phase noise. Moreover, it appears to be the natural choice in zero-g environments where the atoms are initially at rest with respect to the retro reflection mirror. In this case, no Doppler effect lifts the degeneracy between the two possible beam pairs driving a Raman transition. Consequently, both beam pairs interact with the atoms and cause a symmetric beam splitting which transferes both interferometer arms into the same internal state. All three laser pulses forming the Mach-Zehnder like $\pi/2 - \pi - \pi/2$ interferometer are applied by the same optical setup. A polarization maintaining optical fiber guides both the $^{87}$Rb and $^{85}$Rb beam splitter light fields to the science chamber (see Figure 6). Both are pulsed onto the two atomic species at the same time. After the fiber the beams diverge due to the numerical aperture of the fiber plug. A single lens then collimates the beams with a $1/e^2$ radius of 2 cm. Inside the vacuum chamber, a mirror retro-reflects the beam splitter light fields into themselves, thus creating their counter propagation. Placing the mirror inside the vacuum system avoids disturbances of an additional view port in the retro reflection path. In front of the mirror a λ/4 retardation wave plate is mounted. The interferometry sequence consists of three laser pulses separated by the free evolution time T=5 s. First, a π/2-pulse with duration of 50 µs coherently splits the atomic ensembles into a superposition of two momentum states but same internal state. Then a π-pulse redirects the movement, changing the external state, but again keeping the internal state. Finally, a π/2-pulse recombines the trajectories. During the interferometry, the offset 2 coil pair lifts the degeneracy of the magnetic sub states by applying an offset field of 1 mG.

While the atoms propagate on both upper and lower trajectory in the same internal state during the free evolution time the output ports differ in the internal state (see Figure 1, output ports depicted in green and blue).

The retro reflector is a mirror with a diameter of 5 cm. A peak to valley surface quality of λ/50 and surface roughness < 2 nm rms are required to meet the target accuracy. Flat optical mirrors with full aperture up to 32 inch meeting these requirements are manufactured by ZYGO [41]. The mirror is mounted inside the vacuum chamber on a remote controllable mirror holder with a required tuning range of few mrad and a required resolution of ~1 µrad. Current close loop steering mirrors with a sufficient range of 3 mrad achieve an angular resolution of 1.6 µrad and were qualified for the PHARAO/ACES mission [42]. The proposed design might require changes to support a larger mirror, to comply with outgassing restrictions inside the vacuum chamber, and the magnetic cleanliness requirements. The capability to tune the mirror orientation in two axes allows for optimizing the overlap of the incoming and reflected beam. Actuation is not used during nominal science operation phase.

## 5.3 Detection System

The detection procedure aims to fulfil a shot-noise limited detection for $10^6$ atoms in each species assuming the expansion parameters at Thomas-Fermi radii of 4 mm after 10 s of expansion. The detection noise must be well below the atomic shot-noise level (1/3) for the required final precision not to be affected strongly. Moreover, overlapping clouds of $^{85}$Rb and $^{87}$Rb need to be detected at the output ports of two simultaneous double-diffraction interferometers. Systematic effects could appear in the detection of the 4 states (2 species, with 2 states each) which should be understood and suppressed where possible.

In Table 5, the required photon number scattering from a detection pulse is given, both for fluorescence and absorption scheme such that the atom shot noise is dominant by a factor of 3 over the photon shot noise at the detector [43]. Due to the inferred large number of pixels needed for the CCD in absorption imaging to reach atom shot noise (see Table 5), this method is not preferred. However, absorption detection would still be feasible if the detector would be a photodiode or if the cloud would be compressed before detection, leading to a higher proportion of photons scattered and thus a higher SNR with fewer photons. The desirable advantage of absorption imaging is its simplicity and robustness against misalignments.

Therefore as baseline, fluorescence imaging is foreseen during nominal operation for the simultaneous detection of all three interferometer exit ports while absorption imaging will be used during the calibration procedures for source optimization and test measurements.

**Table 5:** Characteristic parameters for fluorescence and absorption detection.

| Parameters | Fluorescence | Absorption |
|---|---|---|
| Number of detected photons to reach target SNR | $9*10^6$ | $\sim 5.4 \times 10^{11}$ (number of photons passing through the atom cloud) |
| Pulse time | ~1 ms | 1ms |
| Total power | 2 mW | < 2 mW (to stay below saturation) |
| Minimum CCD pixel number onto which the atom cloud is imaged, assuming a pixel well depth of $10^5$ electrons | > 200 | > 5 M |

The minimum photon number was evaluated using a numerical aperture of 0.17 corresponding to the lens imaging the atom clouds ($NA_1$ in Figure 8). This value is constrained by the dimensions of the vacuum chamber and an assumed cloud separation of 12 mm. To satisfy our shot noise requirements and be able to image both clouds simultaneously on a CCD we require an image demagnification of 3.

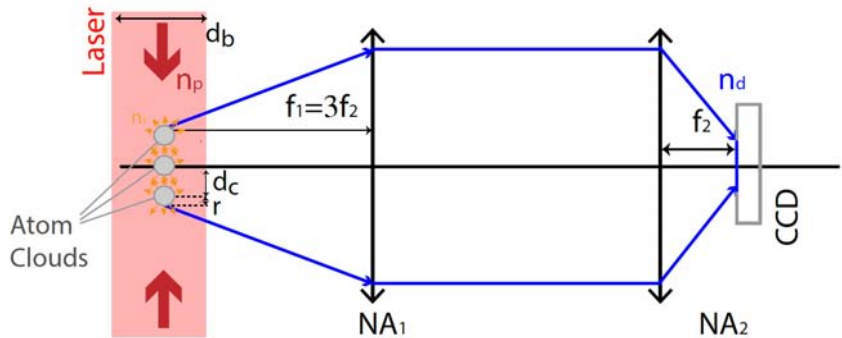

**Figure 8:** Sketch representing the fluorescence detection scheme: the blue line is the fluorescence beam to the photodiode, while the red beam is the probe laser beam.

As defined in [43], simultaneous fluorescence imaging of the two states for each species using a CCD is chosen to reduce the requirements on laser noise and a re-pumper beam is included during fluorescence to avoid atom loss to non-closed hyperfine transitions. The different atom species are imaged in successive sequence to avoid spontaneous emission of one type of atoms when detecting the other.

The suggested laser parameters for detection of 4 mW total power with a 40 mm $1/e^2$ beam diameter for an intensity of $I_0 = 0.32$ mW/cm$^2$ put the intensity seen by the atoms below the saturation intensity (1.67 mW/cm$^2$). For illumination of the atomic clouds, the light fields forming the Raman beam splitters are adjusted in frequency and power.

Considering the effect of 3 MHz detuning and of the chosen saturation intensity, the necessary minimum detection time is 730 µs (see [43]). The suggested pulse time of 1.5 ms thus has a sufficiently high "safety margin" to compensate other noise sources, e.g. in the readout electronics. It will not affect background light originating from the detection laser itself, which has to be reduced by other measures, such as blacking of as many parts of the apparatus as possible (at our parameters less than 1 in a million photons can be allowed to scatter into the detector by means other than the atoms themselves).

The fluorescence detection sequence is detailed in Table 6.

**Table 6:** Detection sequence. The output of the two species is detected sequentially, while the output ports of the individual species are detected simultaneously.

| Time | Laser frequency | Duration |
| --- | --- | --- |
| 0 ms | Detuned by 3 MHz below the F=2 → F'=3 $^{87}$Rb transition and on resonance with the F=1 → F'=2 $^{87}$Rb transition (repumper) | 1.5 ms (730 μs min) |
| 1.5 ms | Shift of image to protected CCD area | 2 ms |
| 3.5 ms | Detuned by 3 MHz below the F=3 → F'=4 $^{85}$Rb transition and on resonance with the F=2 → F'=3 $^{85}$Rb transition (repumper) | 1.5 ms (730 μs min) |
| 5 ms | Shift of image to protected CCD area | 2 ms |
| 7 ms | CCD readout | 2 s |
| 2007 ms | Pusher beam: on resonance with the F=2 → F'=3 $^{87}$Rb transition, and on resonance with the F=3 → F'=4 $^{85}$Rb transition | 10 ms |
| 2017 ms | Background image 1: Detuned by 3 MHz below the F=2 → F'=3 $^{87}$Rb transition | 1.5 ms (730 μs min) |
| 2018.5 ms | Shift of image to protected CCD area | 2 ms |
| 2020.5 ms | Background image 2: Detuned by 3 MHz below the F=3 → F'=4 $^{85}$Rb transition | 1.5 ms (730 μs min) |
| 2022 ms | Shift of image to CCD area | 2 ms |
| 2024 ms | CCD readout | 2 s |

The noise requirements for the camera have been derived [43] assuming a back illuminated CCD chip with a pixel well depth corresponding to $5 \times 10^4$ photons detected per pixel and a quantum efficiency of η = 0.7. With a readout noise of $\sigma_{CCD} \sim 5e^-$ we require to have more than 200, but less than $1.74 \times 10^5$ pixels. For a CCD size of 8.8 mm x 13.3 mm this corresponds to a pixel size larger than 25 μm$^2$.

During detection we must account for the systematic error due to the cross fluorescence signal, which arises from the fluorescence image induced on one isotope from the detection beam tuned to the detection resonance transition of another isotope. The ratio between the scattered (and the detected) photons induced on the F=3 → F'=4 transition ($5^2S_{1/2} \rightarrow 5^2P_{3/2}$) in $^{85}$Rb by the detection beam which is 3 MHz detuned with respect to the one scattered on the F=2 → F'=3 transition ($5^2S_{1/2} \rightarrow 5^2P_{3/2}$) in $^{87}$Rb, is given by the equation:

$$\frac{(n_s)_{Rb85}}{(n_s)_{Rb87}} = \frac{1 + 4\frac{\Delta_{Rb87}^2}{\Gamma^2} + \frac{I}{I_{sat}}}{1 + 4\frac{\Delta_{Rb85}^2}{\Gamma^2} + \frac{I}{I_{sat}}}$$

with $\Delta_{Rb85}$ = 1.1 GHz the detuning of the $^{87}$Rb detection beam with respect to the considered $^{85}$Rb detection transition, $\Delta_{Rb87}$ = 3 MHz, and where we assumed an equal number of $^{85}$Rb and $^{87}$Rb detected for each state cloud. Under these conditions, we find $(n_s)_{Rb85}/(n_s)_{Rb87} = 3.4 \times 10^{-4}$ which induces a detection systematic error near the atom shot noise error. Therefore it will be necessary to introduce an additional calibration procedure to remove this cross isotope readout systematic error.

## 5.4 Vacuum Pumps and Vacuum Technology

In order to conduct atom interferometry experiments, an ultra-high vacuum in the range of 10$^{-11}$ mbar is needed. Only few types of pumps are able to maintain such an environment. Two pump types will be used in the vacuum system: An ion getter pump is mainly used to pump noble gases and methane, while a passive getter pump is used to pump reactive gases like water and hydrogen. Due to the magnetic field created by the ion getter pump it has to be placed outside of the magnetic shield. Both types of pumps have been used on ACES/PHARAO (with a required vacuum of $2\cdot10^{-10}$ mbar) [35][36] and an adapted design based on these experiences will be used for the proposed setup.

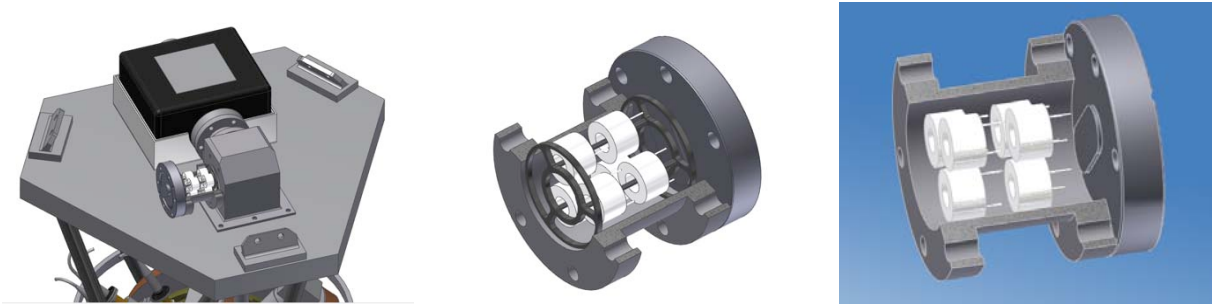

**Figure 9:** Vacuum pumps. On the left side, the whole pumping section consisting of an ion getter pump (black) and a passive getter section is depicted. The six passive getters are mounted inside a small tube (middle, right).

In the current design (see Figure 9) the following parts are foreseen: six passive getter pumps of the type SAES ST171-NP-16-10 and an ion getter pump (IGP) of the type Agilent Technologies VacIon Plus 20 with Sm-Co magnets. The six getters are housed in a titanium tube that includes the electrical interfaces for getter activation. The chamber will be build out of grade 5 Titanium (Ti-6Al-4V). This material is chosen because of its excellent magnetic properties: It has a magnetic susceptibility of $4.1\ 10^{-5}$ (SI units) which is approximately 1000 times smaller than that of Stainless Steel 316L (X2CrNiMo17133E). The Titanium alloy is also superior in its mechanical properties especially with respect to tensile strength and density.

All viewport glasses on the experiment consist of anti-reflex coated BK-7 substrates in order to assure minimal loss. BK7 has the advantage of an almost matching thermal expansion coefficient with titanium.

For the performance of the vacuum system, a careful choice of the seals is essential. There are three different types of seals in the setup. For metal-metal connection on the central part of the vacuum system, diffusion brazing will be used: The pieces are heated to about 80% of the temperature of the melting point and then a force is applied pushing the pieces together. The created joints have strength and leak tightness comparable to a part machined of the bulk material and have the advantage of sizes compared to CF-seals. For metal-metal connections in the peripheral parts of the vacuum system CF-seals according to ISO 1609 will be used. These are sufficient if enough space is available. For metal-glass seals Indium-Lead brazing will be used.

## 5.5 Magnetic Shielding

Magnetic stray fields have to be suppressed, since magnetic field gradients induce differential accelerations between the two Rubidium isotopes. The requirements are to keep gradients below 500 µG/m over the baseline of the atom interferometer during the atom interferometer pulse sequence and below 3 µG/m at the location of the ODT for several 10 ms after release of the atomic ensembles.

The selected baseline for magnetic shielding is to use 4 shielding layers, each shield 1 mm thick with gaps between subsequent shields of 13 mm in radial and 35 mm in axial direction, cf. Figure 4. Each shield is assembled from two segments. The shield configuration provides various feedthroughs for a vacuum tube connecting the pump section to the science chamber, for heat strap connection from the science chamber to the middle plane, and for feedthrough of optical fibers and electrical cables. The top segments provide 3 more openings for mounting the science chamber to the middle plane. The key parameters of the proposed magnetic shielding are given in Table 7. As observed in former experiments, the realized magnetic shielding factor is usually lower (up to a factor of 4) than in the corresponding FE simulation [44].

The mounting of the shield will be done such that any mechanical stress on the shield is avoided. The supporting struts for the vacuum chamber system shall thus be connected to the middle plane of the payload platform through the 3 openings in the top segments of the µ-metal shield. No vacuum chamber mounting struts are attached to the µ-metal shields themselves.

**Table 7:** Mass and outer dimensions of the magnetic shielding and results from FE-simulations for shielding factors S along principal axes.

|  | Magnetic shielding configuration |
|---|---|
| mass | 50,7 kg (no harness) |
| outer dimensions | h = 798 mm, d = 536 mm |
| $S_x$ | 140 222 |
| $S_y$ | 141 106 |
| $S_z$ | 41 125 |
| no. of layers | 4 |
| layer thickness | 1/1/1/1 mm |
| gap size | radial: 13mm / axial: 35 mm |

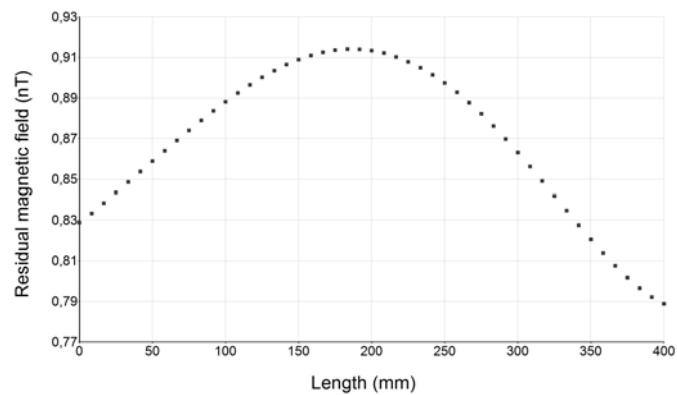

**Figure 10:** FE simulation of magnetic field gradients from external fields (external B-field of 40 µT along the cylinder axis; resulting gradient along the cylinder axis: <0.06nT/15cm). Shown in the graph is the residual field (in nT) over length (in mm) [44].

Finite-element simulations for various external fields show the magnetic field gradient along the cylinder axis to be less than 0.1 nT/12 cm (see Figure 10). Dominating field inhomogeneties will be due to residual magnetization of the shields though and not due to residual external fields. It is expected that field variations due to residual magnetization can be kept below 1 nT/12 cm. The effect of such a field gradient with T = 5 s and $B_0$ = 100 nT leads to a phase shift of $< 1 \cdot 10^{-12}$ m/s² (factor 500 above requirements). This potential obstacle will be met by a careful calibration of the constant residual fields inside the magnetic shield and by application of an interferometer sequence with input states that are subsequently alternated between hyperfine levels of the ground state. During each evolution, 0.25 h around apogee will be used to assess the bias due to magnetic field gradients. The strength of the offset field during interferometry will be increased compared to nominal operation at perigee to different values which also linearly increases the bias and subsequently allows the estimation of the gradient. Thus a cancellation of the phase shift due to magnetic field gradients on the order of a factor > 500 shall be achieved.

In addition to passive shielding, the baseline design also foresees active compensation of slowly varying external fields (> 1 s) similarly done in PHARAO. To this end, the middle shield shall be equipped with magnetic coils and probes to monitor the magnetic field allowing the application of a compensation field within an active feedback loop. For magnetic field monitoring, anisotropic magnetoresistors (AMR) are foreseen, based on experience with LISA Pathfinder and further research for LISA [45]. They offer smaller size and lower power consumption than conventional fluxgate magnetometers used in space missions. Moreover, the lower magnetization of the AMR allows placing the sensor closer to the region where magnetic field needs to be measured and, in consequence, a more accurate map reconstruction of the magnetic field can be made [46].

Exposure of the µ-metal to magnetic fields of the order of mT and above, of either external or internal origin, causes residual magnetization of the µmetal, thus deteriorating the shielding factor. Elimination of this residual magnetization requires degaussing of the shield during commissioning and potentially at regular intervals during the course of the mission. For this purpose each shield is provided with a degaussing coil of 10 windings of copper wire (summing up to 8 degaussing coils in total). The foreseen procedure is to apply a slowly oscillating current (0.1 – 2 Hz) of up to 10 A, which will be ramped down to zero amplitude within few seconds.

## 5.6 Thermal Control System

One of the critical issues is the avoidance of position and length changes due to thermal expansion. These changes have impacts on the magnetic field, which is critical for the sensor operation. Consequently, arising from the need for a magnetic field gradient of dB/dz < 3 µG/m during the preparation sequence, maximum allowed thermal expansions of the Feshbach coils can be formulated as listed in Table 8 [44].

**Table 8:** Thermal requirements and FE-model results.

| Instrument requirements | Thermal requirements | Simulation results |
|---|---|---|
| Change of the distance between the Feshbach coils: $\Delta l < 320$ nm | Temperature stability of the vacuum chamber for 2000 s: $\Delta T < 0.30$ K | $\Delta T < (0.014 \pm 0.004)$ K |
| Change of the Feshbach coils diameters: $\Delta d < 520$ nm | Temperature stability of the Feshbach coils for 2000s: $\Delta T < 0.15$ K | $\Delta T < (0.11 \pm 0.03)$ K |

The design approach for the thermal control system (TCS) of a space compatible atom interferometer is challenging, since constraints resulting from the design of the interferometer itself, as well as constraints arising from the satellite mission and the satellite bus have to be considered. In orbit thermal environmental conditions may change rapidly and are much more unstable compared to laboratory conditions. In this study we consequently consider thermal boundaries and requirements as derived by the STE-QUEST mission [17].

The high amounts of waste heat produced by the electronics and magnetic coils have to be transported to the satellite's heat sink. In order to create a robust and effective system, the TCS is designed as a complete passive system. The main challenge is the management of the cyclical instrument operation and the resulting fluctuating heat fluxes. In order to develop a TCS, which fulfills these requirements, a finite-element (FE) model is created for validation on its thermal stability. The analysis is performed with the FE-software ANSYS Classic [47]. Due to the high complexity of the atom interferometer physics package, the FE-model can only be an idealized representation of the physics package. Nevertheless the geometry diameters and the mass of the components are equivalent to the AI design, due to their thermal importance as heat capacities and thermal conductors.

Thermal radiation to the magnetic shielding is neglected in this model as well as the holding structure. Both would lead to a better thermal transportation and cooling of the critical components, which would even improve the performance of the TCS. In this sense the resulting TCS is designed with respect to a thermal worst case scenario.

Table 9 shows the operation times of the components of the physics package within the 20 s experiments cycle and their heat impacts. The most critical property is the very high waste heat produced by the Feshbach coils, especially since they are mounted directly on the vacuum chamber. The time-varying operation of the other components is critical as well and has to be buffered to ensure stable temperatures.

**Table 9:** Components heat dissipation of one experiment cycle of 20 s included in the FE-model. Mesoscopic U, Mesoscopic H, Base-chip and Science-chip are all realized as part of the atom chip.

| Component | Operation Time [s] | Peak Power [W] | Averaged Power [W] |
|---|---|---|---|
| Mesoscopic U | 0-2 | 22.5 | 2.25 |
| 2DMOT | 0-2 | 13 | 1.3 |
| Offset 1 | 0-2 | 2x32.5 | 2x3.25 |
| Mesoscopic H | 2-2.1 | 50 | 0.25 |
| Base-chip | 2-2.1 | 50 | 0.25 |
| Science-chip | 2-5.6 | 8 | 1.44 |
| Feshbach Coils | 5.6-8.9 | 2x67.5 | 2x11.1375 |
| CCD Camera[1] | 10-20 | 5 | 2.5 |
| Dispenser Heater | 0-20 | Constant on 40°C | Constant on 40°C |

---

[1] In the FE-model only one CCD camera is included, while three will be necessary for the final design.

The TCS design of the physics package is shown in Figure 11. The goal of the design is to achieve a very high thermal stability with consideration of the time-varying instruments heat dissipation and changing environmental condition. The instrument's bus heat sink is the middle plate, which temperature needs to be stabilized to 13°C ± 3°C.

In Figure 11, the thermal connections between the components are represented by high conductive carbon fiber heat [48] strap (black line), thermal insulation (red line) and mounting connection (purple line). High conductive heat straps with individually calculated diameter sizes couple the heat sources to a cold plate, which is a mounting, heat storage and heat exchanging unit. This will ensure high waste heat transportation rate out of the system while the time-varying operation and environmental changing can be buffered by the cold plate. As the atom chip is mounted inside the vacuum chamber, a thermal connection is established through a copper gasket.

To avoid heat impacts to the thermal sensitive components, thermal insulation is installed in between. The thermal properties of the mounting connections have to be considered as well and are included in the FE-model.

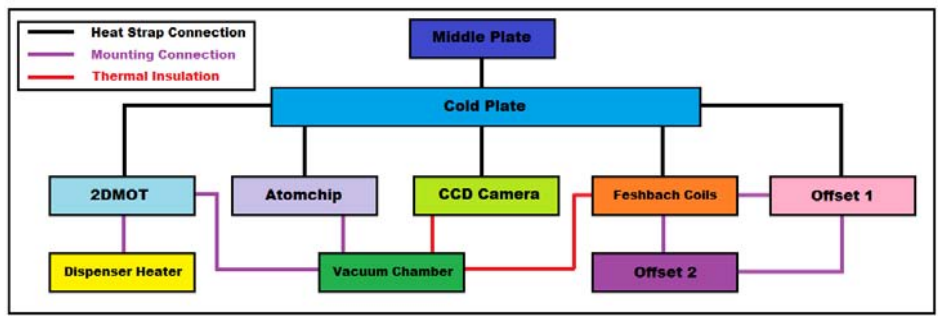

**Figure 11:** Schematics of the physics package TCS design.

The FE-model representation is shown in Figure 12. The magnetic coils are made of copper wires with a reduced thermal conductivity (80 W/(m·K)), resulting from the wire insulation material. The coils are held on aluminum coil holders, which are also used as mounting interface of the heat straps.

A cold plate consisting of high thermal conductive copper is mounted on the top and has a thermal connection to the middle plate by a heat strap of 3.5 cm diameter size. The telescope for the CCD camera is made of aluminum and has a thermal insulation between the camera and the telescope. The dispenser heaters consist of porous titanium. In order to improve the thermal connection to the atom chip through the vacuum chamber, a copper gasket is applied as heat strap interface on the outside of the vacuum chamber. In addition to the model displayed in Figure 12, heat straps in terms of direct node to node 3D-Link elements and thermal insulation between the Feshbach coils and the vacuum chamber are included in the simulations.

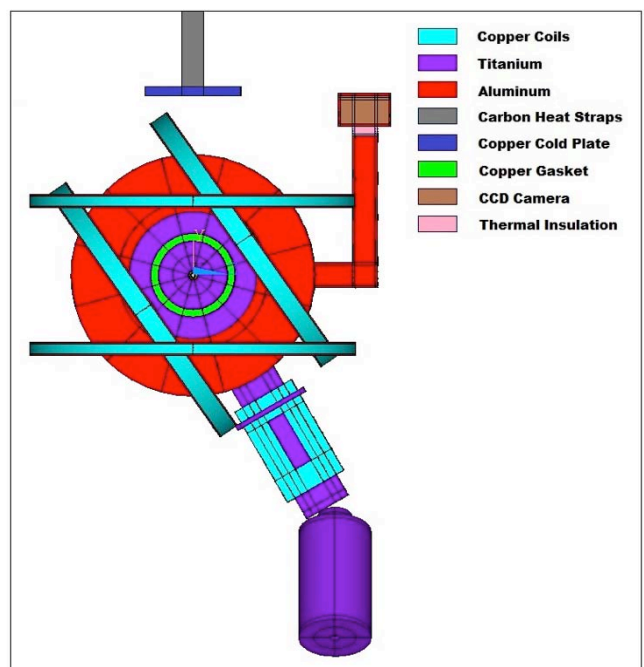

**Figure 12:** FE-model of the physics package and the model materials.

The TCS has to be thermalized to the system's equilibrium temperature within the calibration time of the mission [44]. Once it has reached the thermal equilibrium, the results of the FE simulations show that the thermal requirements are fulfilled by the TCS (see Table 8).

Figure 13 shows the time evolution of the temperatures of the components for 1000 seconds and under environmental changing condition (± 3°C middle plate temperature change every 20 seconds). While the atom chip and the cold plate are showing cyclical behavior, the vacuum chamber, the Feshbach coils and 2DMOT coils are stable. The changing environmental temperature has only small effects on the critical components as well.

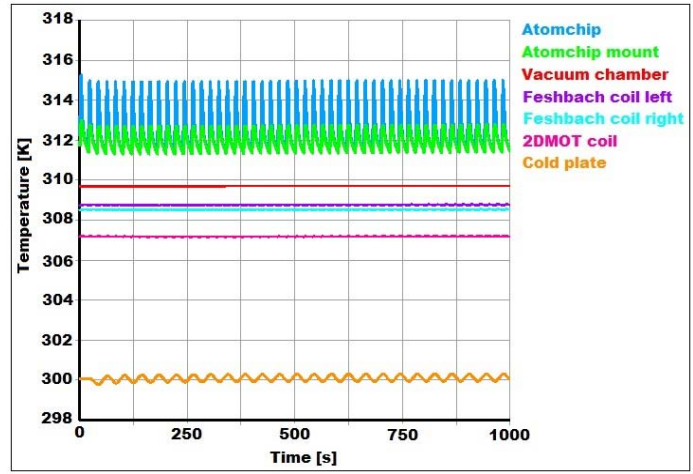

**Figure 13:** Time evolution of the components temperatures during 1000 seconds.

After the diameter size validation of each individual heat strap with the FE-model, a CAD design of the physics package with the thermal straps is created as depicted in Figure 14.

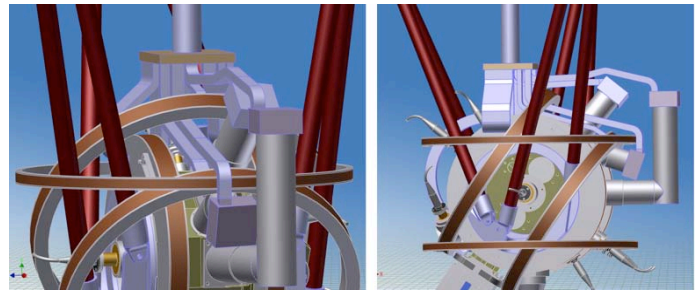

**Figure 14:** CAD model of the physics package including the holding structure and the TCS.

## 5.7 Physics Package Technology Readiness Estimation

The Physics Package design presented in this chapter is largely based on work carried out within the drop-tower experiment QUANTUS and the sounding-rocket mission MAIUS, both performing BEC-based dual species atom interferometry using Rb and K atoms. This includes optimization of the instrument with respect to compactness, mechanical stability and autonomous operation. The MAIUS development further includes vibrational testing but no thermal cycling and no radiation hardness tests. With respect to space application as discussed in this paper, a technology readines level (TRL) between 4 and 5 is therefore assigned to most components of the Physics Package, as no specific tests under relevant environment are yet carried out. Dual species atom interferometers using $^{87}$Rb and $^{85}$Rb were already operated [7][52], but for STE-QUEST additional atom optical techniques have to be considered. Their combination remains to be demonstrated. A TRL between 3 and 4 is assigned to the science chamber as similar chambers are tested in ACES, MAIUS, but a redesign and a different thermal management is necessary. The atom chip needs a redesign with microwave antenna, resulting in TRL3. TRL 3 is also assigned to the dispenser, whose design is based on the development within the PHARAO project but needs adaption from Cs to Rb, and to the retroreflector actuator which is also based on PHARAO heritage but needs revision for higher loads.

# 6. Laser System Design

Precise atom interferometric measurements of the relative acceleration between degenerate quantum gas mixtures of $^{87}$Rb/ $^{85}$Rb require an advanced laser system which is capable for (i) simultaneous laser cooling of the atomic species as well as internal state preparation, (ii) all-optical two-species BEC generation, (iii) coherent matter wave manipulation by means of symmetric two-photon Raman transitions, and (iv) successive detection of the output states of the two scaling-factor matched interferometers.

Therefore, a hybrid system comprising a reference and optical dipole trap laser (AI-ROL) based on telecom technology and frequency doubling techniques with micro-integrated, high power diode laser modules (AI-DLP) is foreseen. The complex switching procedures of all laser beams according to the experimental sequence as well as the precise and controlled distribution of the laser light to the physics package (AI-PPU) is realized with a Zerodur optical bench setup combined with a fiber optical splitter system (AI-SDM). This hybrid concept combines major advantages of the two technologies, yielding to a comparatively high overall TRL and allowing for acceptable budgets with respect to power consumption and mass by providing all required functionalities.

Each of the three laser subsystems (ROL, DLP and SDM) is integrated within a separate mechanical housing structure optimized with respect to thermal budgets and adapted for singularities, e.g., spatial characteristics of key components. A functional scheme of the laser system design is given in Figure 15.

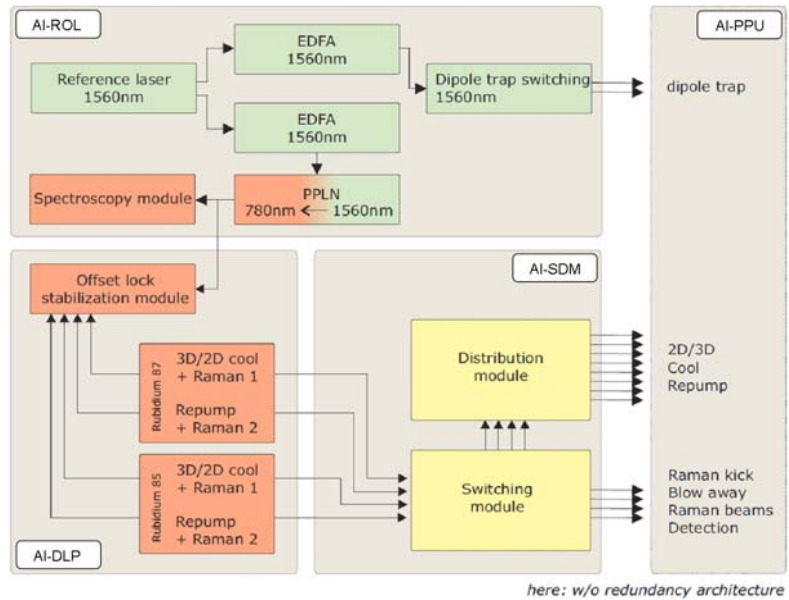

**Figure 15:** Schematic of the laser system for dual-species atom interferometry, which combines a telecom technology based reference and optical dipole trap laser module (AI-ROL) with micro-integrated high power diode laser modules (AI-DLP). The latter is used for performing simultaneous cooling in a 2D$^+$/3D MOT configuration, internal state preparation, coherent manipulation and detection of $^{87}$Rb and $^{85}$Rb quantum gas mixtures. For switching, combining and distribution, the light of 4 diode laser modules is first guided to one consolidated switching module and afterwards to a distribution module (AI-SDM), both being directly connected to the physics package (AI-PPU). All interfaces shown here are polarization maintaining (pm) single mode optical fibers**.**

## 6.1 Reference and Optical Dipole Trap Laser Module

The Reference and Optical Dipole Trap Laser Module (AI-ROL) has to support the generation of an absolute frequency reference at 780nm by stabilizing a reference laser with respect to the $|F = 3> \rightarrow |F = CO\ 3/4>$ transition of the D2 line of $^{85}$Rb (0.1 MHz linewidth, 30 mW output power and better than $3\cdot10^{-10}$ long term frequencz stability). Moreover, the AI-ROL delivers the light for forming the crossed all-optical dipole trap in the vicinity of the atom chip (see Chapter 5.2). In both fibers together, the dipole trap should deliver 2 x 1 W of $\lambda = 1560$ nm ($\Delta \lambda < 1$ nm) light to the atoms, with a power stability of less than 0.1 %.

The AI-ROL includes the frequency stabilized reference laser serving as master laser for the diode lasers which are part of AI-DLP and generate the laser beams for optical dipole trap operation. A schematic of this module is given in Figure 16.

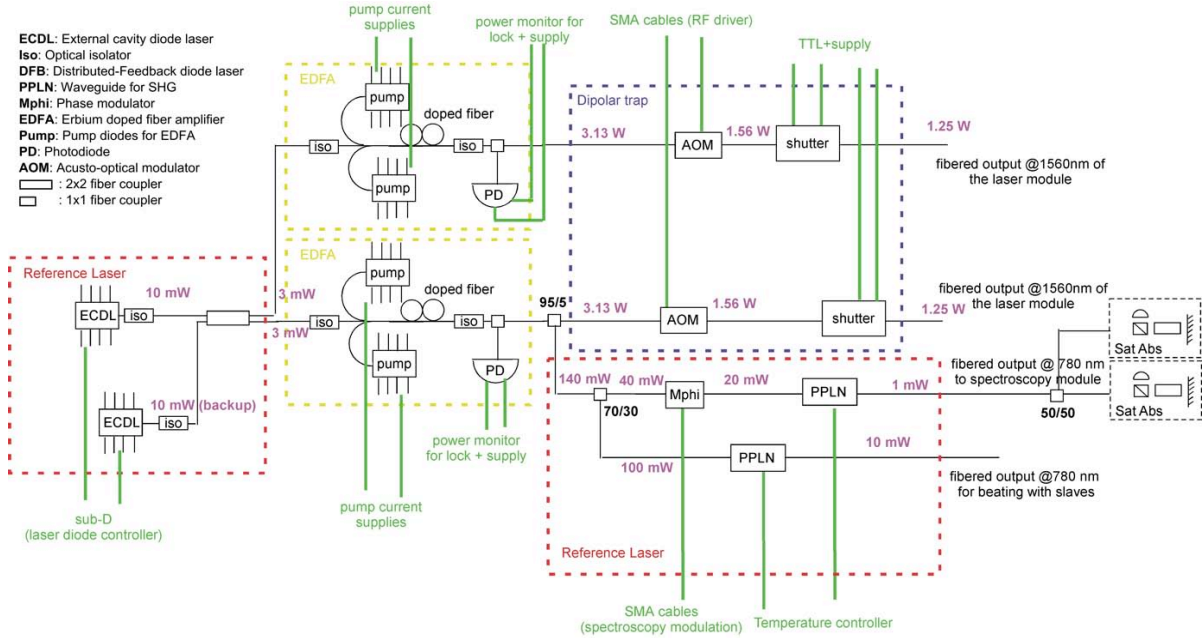

**Figure 16:** Schematic of the AI-ROL subsystem architecture including redundancy. It consists of two external cavity diode lasers (ECDL), which are overlapped with in a 50/50 fiber splitter. Two fibered amplifiers (EDFA) are used forming the two beams of a crossed optical dipole trap. Each EDFA features two sets of pumping diodes for redundancy purposes. Light from one path is additionally used to generate an optical reference at 780 nm by means of frequency doubling techniques. An additional 50/50 fiber splitter @ 780 nm is implemented after the second PPLN, delivering the light to the two spectroscopy modules.

The AI-ROL is based on all-fibered telecommunication components and frequency doubling waveguide technologies. The relevance of this choice is an outcome of the phase A study. Technological maturity has been successfully demonstrated within important projects in the area of inertial quantum sensors: (i) The I.C.E. experiment towards a quantum matter based test of the weak equivalence principle in microgravity with a Rb/K interferometer and (ii) MINIATOM, whose goal is to build a compact and transportable gravimeter with cold atoms. Suitability of the main components of the setup for space applications has been determined in [49].

The all-fibered components developed in the telecom field naturally supply miniaturized, compact and extremely robust solutions. The master laser is a very narrow linewidth (10 kHz) external cavity diode laser (ECDL), emitting a few mW at 1560 nm and housed in a butterfly package (Redfern Integrated Optics). After passing a fibered isolator, this telcordia qualified laser module injects two efficient Erbium doped fiber amplifiers (EDFA), optically pumped with high power laser diodes at 980 nm to generate sufficient amplification at relatively small length scales. Both outputs of the EDFAs are used to generate two single beams forming a crossed optical dipole trap. Besides power locking capabilities and sufficient optical isolation, each arm features a single fibered acousto-optical modulator (AOM) to switch the beams of the crossed dipole trap independently. AOMs are also used to reduce the optical intensity in both of the arms for systematic evaporation and to realize precisely controlled optical pulses for dual-species delta-kick cooling [33].

After one EDFA output, a fibered, polarization maintaining (pm) beam splitter guides approximately 1% of the 1560 nm light in a fiber-based phase modulator and afterwards through a periodically poled Lithium-Niobate (PPLN) waveguide for non-linear frequency conversion. This step is necessary to generate the reference light at 780 nm for rubidium spectroscopy. Doppler-free frequency modulation spectroscopy (FMS) is realized in a rubidium vapor cell setup based on Zerodur bonding technology [53].

In this spectroscopy module, the 780 nm light coming from an optical fiber is collimated by a glass-ceramic based fiber coupler and split into two paths using free space optics integrated upon the optical bench with adhesive bonding techniques. For realizing a robust frequency stabilization with the capability of an automated relock, two signals will be recorded. The Doppler-broadened signal is generated by detecting a beam which passed the vapor cell once. This signal is utilized for gaining an initial value for an automated relocking system. The Doppler-free signal is generated by retroreflecting the beam after one passage through the vapor cell and detecting it after the second passage by a shot-noise limited photodiode assembly. We anticipate a long-term frequency stability of better than $3 \cdot 10^{-10}$ by stabilizing the laser with respect to the $|F = 3\rangle \rightarrow |F = CO\ 3/4\rangle$ transition of the $D_2$ line of $^{85}$Rb.

As redundancy option, the spectroscopy module is implemented twice into the setup (see Figure 16), providing the option to electronically switch between the error signals of both spectroscopy units for stabilization of the master laser and advanced automated re-locking techniques.

A separate fibered, low-power PPLN generates 780 nm light output, which is guided towards a fiber-optical splitter system to distribute and overlap the frequency-doubled reference laser light with the light of four microintegrated diode laser modules (see Chapter 6.2). The generated beat notes, typically in the GHz regime, are detected with fast photodiodes and subsequently used for frequency stabilization of the four diode laser modules.

A prototype of the whole system is currently realized in the laboratory to validate the strategy and the architecture of the instrument. For this purpose, a commercial version of this laser has been built by muQuanS. The issue of power consumption is not taken into account for this first version.

The AI-ROL module with housing is shown in Figure 17, with dimensions of 310 x 310 x 100 mm$^3$, a total mass of 7.1 kg (8.52 kg incl. 20% component level margin), a total average power consumption of 35.46 W (42.55 W incl. 20% component level margin) and a peak power of 52.20 W (62.64 W incl. 20% component level margin).

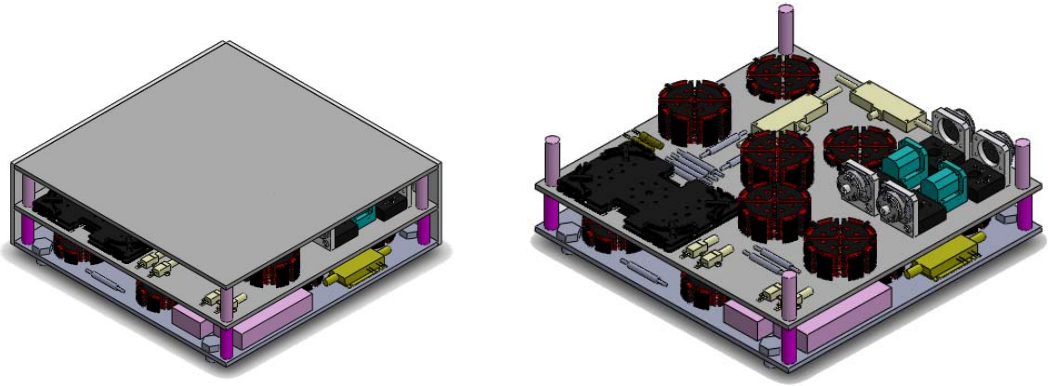

**Figure 17:** Envisioned two level design of the integrated AI-ROL with external dimensions of 310 x 310 x 100 mm³. (c) University of Bordeaux.

## 6.2 Diode Laser Package Module

The sources for laser cooling (in a 2D$^+$/3D MOT configuration), internal state preparation, coherent manipulation and detection of $^{87}$Rb and $^{85}$Rb quantum gas mixtures are micro-integrated diode laser modules at a wavelength of 780 nm. For the operation of the dual-species 2D$^+$ and 3D MOT, 200 mW of cooling light and 20 mW of repump light at the atoms position are foressen for each species. With a linewidth of 1 MHz and a long-term stability of better than $3 \cdot 10^{-10}$, a dynamic tuning capabilty of about 200 MHz (< 1 ms) with a relative uncertaintiy of less than 100 kHz need to be ensured. For state preparation and detection (2 x 30 mW), as well as for coherent manipulation with Raman beams (4 x 160 mW), the requirements in linewidth (0.1 MHz), frequency noise (~$7 \cdot 10^5$ Hz$^2$/Hz@ 1 Hz, $10^4$ Hz$^2$/Hz@100 Hz – 10 MHz) and agility (1 GHz within < 2ms with relative uncertainty of < 100 kHz) are more demanding. The long-term power stability should be better than $10^{-2}$ with a RIN of -80dB/Hz @ 1-100 Hz.

For this purpose, a micro-integrated master oscillator, power amplifier concept has been developed to both meet the specific requirements for spectral stability and output power. It features a narrow linewidth, extended cavity diode laser module (µECDL module) as a master oscillator (MO) and a separate high power amplifier (µPA module) unit interconnected with a polarization maintaining single-mode optical fiber. This µECDL-MOPA simultaneously features a narrow linewidth (< 100 kHz) combined with high output power (> 1000 mW). Physical sub-division of µECDL and µPA modules allows for thermal decoupling and efficient thermal management concepts independently optimized for both units, therefore providing improved passive stability.

Figure 18 (left) shows the functional schematic of the optical layout of the µECDL Module (top) and of the µPA Module (bottom). Both systems rely on a basic design consisting of two semiconductor chips, lenses for beam shaping and lensed fiber couplers for the transmission of optical signals in and out of the modules.

In the case of the µECDL Module, the µECDL acting as local oscillator consists of a laser resonator formed between the front facet of a ridge-waveguide (CG1) laser diode and an external volume holographic Bragg grating (VHBG). The output of the µECDL is fed into a GaAs-based phase modulator (CM) that can attain modulation bandwidths in the GHz range. An optical isolator (OD) placed between the ECDL and the phase modulator prevents optical feedback from degrading the frequency stability of the local oscillator. The modulated output of the phase modulator chip is injected into a polarization maintaining single mode optical fiber (OF1) via a lensed fiber coupler (FC1).

The frequency of the ECDL is adjustable via the injection current of the ridge-waveguide diode laser chip and via the temperature of the volume holographic Bragg grating. By both complementary tuning the VHBG and the laser current, µECDLs offer a mode-hop free tuning range of about 40 GHz within a few hundred ms. With an unstabilized linewidth of a few 10 kHz (10 µs) and an intrinsic linewidth of a few kHz, µECDL modules fulfill the key requirements for all and especially for the Raman beams ($7 \cdot 10^5$ Hz$^2$/Hz@ 1 Hz, $10^4$ Hz$^2$/Hz@100 Hz – 10 MHz for frequency noise and 0.1 MHz linewidth).

In the Amplifier Module (µPA), light fed into the module through a polarization maintaining single mode fiber (OF2) is first injected into a RW-chip pre-amplifier (CG2) that provides a gain of at least 10 dB necessary for ensuring saturation of the main amplifier consisting of a tapered amplifier semiconductor chip (CG3). The high power output of the main amplifier is then coupled out of the module via a lensed fiber coupler (FC3) terminated with a polarization-maintaining single-mode fiber (OF3). To reduce influences caused by optical feedback due to unwanted reflections (e.g., non-perfectly AR-coated fiber ends) high-power fiber-optical isolators are integrated after each µPA module.

The diode laser modules are built on either already space qualified or space compatible technologies. The laser chips as well as further optical elements like miniaturized mirrors, micro-lenses, micro-optical isolators and fiber couplers are mounted on a galvanically structured AlN ceramic body. Furthermore, they feature an electronic interface that provides LF (up to a few MHz) current modulation for fast frequency control, a GHz current modulation port as well as miniature temperature sensors for the Aluminum-nitride (AlN) microbench, the laser chips and, where applicable, the volume holographic Bragg grating. All electrical signals are fed into the laser modules via isolated mini-coaxial plugs that support DC signals up to 3 W and HF-signals up to 65 GHz. The laser modules are hermetically sealed in a housing made of Kovar filled with a technical gas and have a size of 128 x 78.2 x 22.5 mm³.

The technological concept is based on the developments in the context of the DLR funded projects QUANTUS, LASUS and MiLas, in which micro-integrated master-oscillator power-amplifier (MOPA) modules have been developed and qualified for sounding rocket missions (8.1 g$_{RMS}$, 20-2000 kHz), featuring GaAs-based DFB master oscillators and tapered amplifiers. The same technological platform has been adapted for the microintegration of ECDLs based on volume holographic Bragg gratings (VHBG). These systems successfully passed 1500 g shock tests and vibration tests with loads of up to 21.4 g$_{RMS}$ within a frequency range of 20-2000 Hz [50][51].

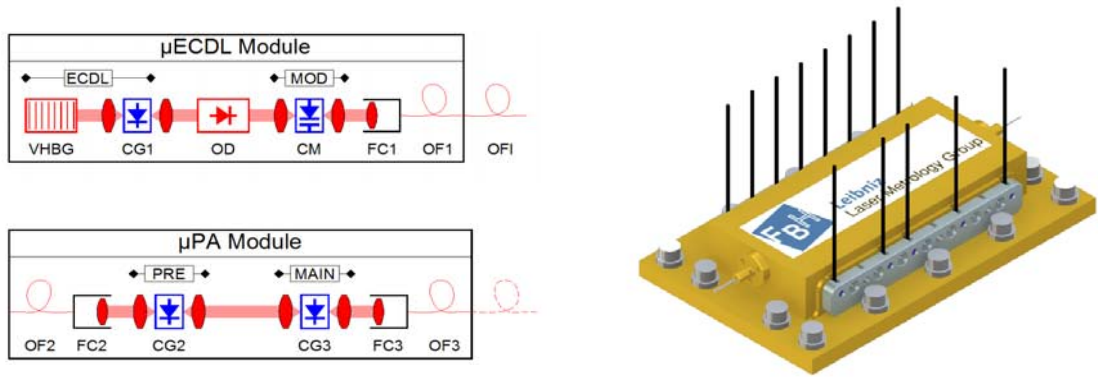

**Figure 18:** Functional interface of the diode laser based µECDL and µPA modules (left) and CAD drawing of the packaged and hermetically sealed modules (right), within a footprint of 128 x 78.2 x 22.5 mm³. (c) Ferdinand-Braun Institut Berlin.

For each species, one µECDL-MOPA system acts as a cooling laser for 2D$^+$/3D MOT operation, which has to be tuned to the $^{87}$Rb ($|F = 2> \rightarrow |F = 3>$) and $^{85}$Rb ($|F = 3> \rightarrow |F ' = 4>$) transition, respectively, with a detuning range from 0 to -20Γ (-120 MHz). Another two µECDL-MOPAs drive the repumping transition of $^{87}$Rb ($|F = 1> \rightarrow |F ' = 2>$) and $^{85}$Rb ($|F = 2> \rightarrow |F ' = 3>$) with relative frequency uncertainties of less than 100kHz.

Preliminary requirements on total optical output power at the atoms are 200 mW of cooling light and 20 mW of repumper for each species with a line width of 1 MHz or less overlapped and distributed onto 4 fibers for the 2D$^+$- and 4 fibers for the 3D-MOT.

After generating the two-species BEC in the crossed optical dipole trap, the aforementioned µECDL-MOPAs are frequency tuned and phase-locked to drive symmetric two-photon Raman transitions between the hyperfine states of $^{87}$Rb and $^{85}$Rb, respectively. Raman transitions are used for controlled momentum transfer within the Raman-kick phase and for all coherent beam splitters in the double-diffraction interferometer. Finally, the same pair of µECDL-MOPAs is utilized for detecting the atomic clouds (incl. blow away procedures).

The requirements of the Raman functionality are stronger in terms of laser linewidth (0.1 MHz) and frequency noise ( $7 \cdot 10^5$ Hz$^2$/Hz @ 1 Hz, $10^4$ Hz²/Hz @ 100Hz - 10 MHz). Their frequencies have to be red detuned between 0 and 1000 MHz with respect to the $^{85}$Rb ($|F = 3> \rightarrow |F' = 2>$ and $|F = 2> \rightarrow |F' = 2>$) and $^{87}$Rb ($|F = 2> \rightarrow |F' = 1>$ and $|F = 1> \rightarrow |F' = 1>$) transitions. Their required power is 4 x 160 mW overlapped in one optical fiber for coherent manipulation and 2 x 30 mW for detection and state preparation. Resonant light resulting from spontaneous emission of the power amplifiers will be suppressed with filter gas cells.

Within a fiber-optical splitter system, the light of all four µECDL-MOPAs is separately overlapped with the master laser (see Chapter 6.1) to generate beat notes for the offset locks, cf. the schematic shown in Figure 19. For redundancy, a functional copy of the four µECDL-MOPA modules, connected to a second offset lock module, is foreseen.

The AI-DLP module with housing is shown in Figure 20. Its dimensions are 400 x 390 x 200 mm$^3$ with a total mass of 22.0 kg (26.4 kg incl. 20% component level margin) and a total average and peak power consumption of 44.40 W (53.28 W incl. 20% component level margin).

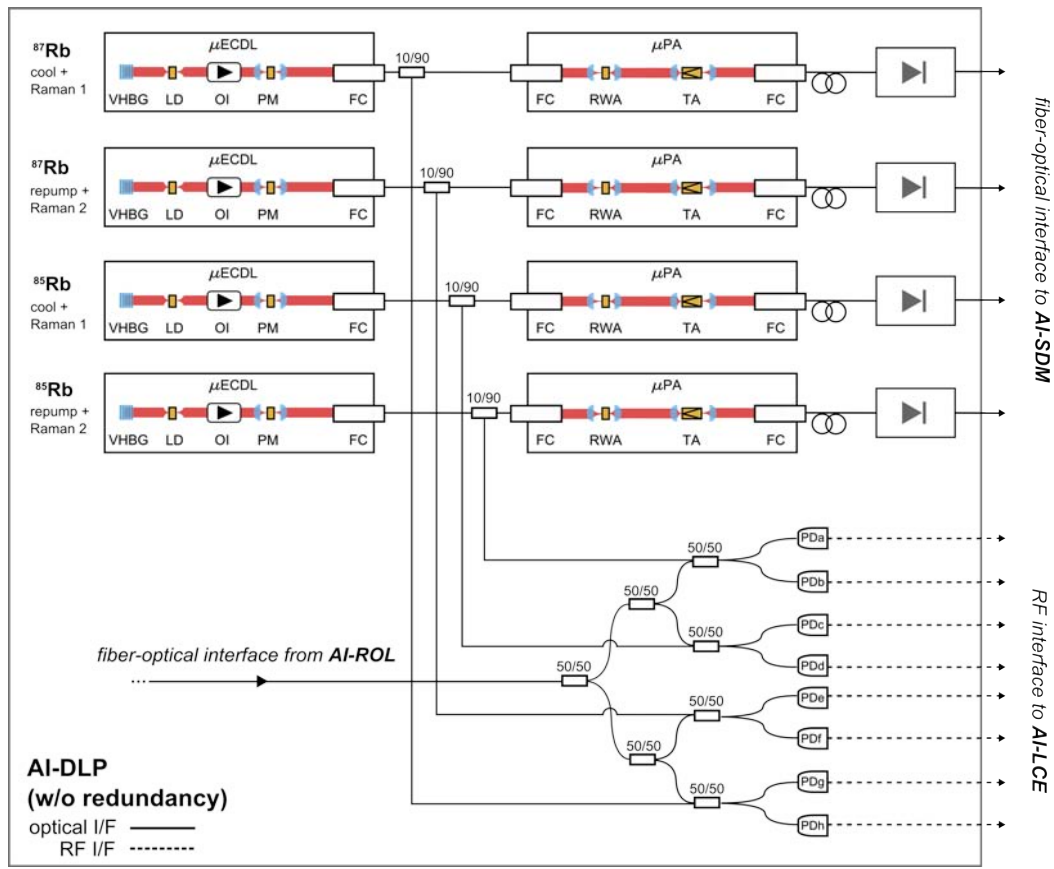

**Figure 19:** Schematic of the AI-DLP subsystem architecture (redundancy is not included, except redundant photodiodes after splitting). Four µECDLs in combination with four µPA modules generate the laser beams for laser cooling, internal state preparation, manipulation and detection. The PA output light is delivered to the Switching and Distribution Module (SDM). Part of each µECDL output is overlapped with light from the Reference Unit for phase locking. The electronic beat signals are processed in the Laser Control Electronics (LCE). (c) Humboldt-Universität zu Berlin.

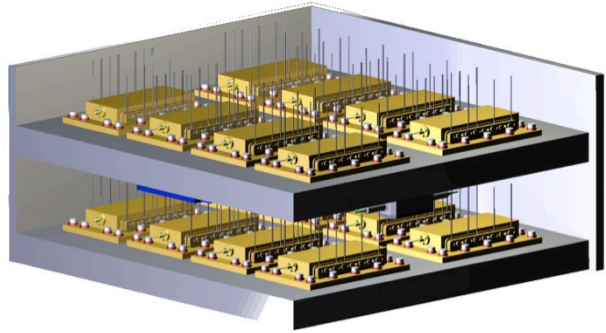

**Figure 20:** Envisioned design of the AI-DLP housing. It consists of a two level structure, where all micro-integrated ECDL modules (μECDL) and the fiber splitters are mounted onto the upper plate, all micro-integrated power amplifier modules (μPA) are mounted onto the lower plate, respectively. Fiber-optical isolators and the offset lock unit are integrated on the bottom side of the top plate. (c) Humboldt-Universität zu Berlin.

## 6.3 Switching and Distribution Module

Precise switching of the laser beams at timescales with adjustable attenuation (30 dB in < 1s, full extinction possible in < 3 ms), power monitoring and user-specified laser pulse generation according to the experimental sequence (cf. Figure 3) is realized on a Zerodur optical bench setup featuring free-space optics and active components which are directly bonded to the zerodur plate [53]. A schematic drawing of the switching module is shown in Figure 21.

The complete module is fiber coupled, so any light is delivered to the board in optical fibers and exits the board in optical fibers as well. The intensity control and switching of the laser beams is realized in a two-stage setup which features a combination of acousto-optical modulators (AOMs) and mechanical shutters. AOMs allow for a precise intensity control by varying the RF-power which is fed into the AOM as well as fast switching times in the low μs-range. Mechanical shutters on the other hand allow for a complete extinction of the laser beams.

After collimation some mWs are split of every beam by a polarising beam splitter and the two frequencies addressing the same isotope are overlapped to allow for a phase lock required for the generation of the Raman pulses. Then each beam passes a shutter-AOM-combination for a frequency selective switching and intensity control. After each AOM the two diffraction orders will be separated by a pick-up mirror, which is positioned such that only one diffraction order hits the mirror surface whereas the other one will miss the mirror completely.

The light in each first diffraction order is used for MOT and molasses operation. After the AOM it passes another mechanical shutter, which allows for an extinction of light in the MOT telescopes during the interferometry phase, and is fed into polarization maintaining optical fibers via Zerodur based fiber couplers. These fibers are connected to the fiber-based distribution module. Each zeroth diffraction order is overlapped by a polarizing beam splitter (PBS). This overlapping results in two beams containing all four frequencies with two being collinear polarized and the other two having orthogonal polarization. One of these two beams is used for Raman interferometry, whereas the other one is utilized for the Raman-kick, detection and blow-away. The interferometry beam passes another AOM required for common-mode pulse generation and intensity control. After the AOM, a few milliwatts are split by a non-polarizing beam splitter, which is used for a primary phase lock. The main beam passes another mechanical shutter and a rubidium gas cell which is required for filtering purposes, as any resonant light affects the interferometer adversely. After passing the cell, the light is coupled into a polarization maintaining fiber, which guides it directly to the physics package. The beam emerging from the second output of the PBS is used for detection, blow away and Raman-kick procedures. From this beam, a small percentage of light is split off to be used for a secondary phase lock. The remaining light passes an AOM and is subsequently split by another PBS. Each beam then passes a shutter and is coupled into a polarization maintaining fiber leading to the physics package. A module using this technology has been built and tested for the MAIUS sounding rocket mission.

The distribution of the four fibers used for MOT operation coming from the switching module to the eight fibers which have to be connected to the vacuum chamber is realized by a solely fiber based splitter array similar to the one used for the offset locking module. Within this module the four different frequencies are overlapped and then split into eight fibers with the required intensity ratio.

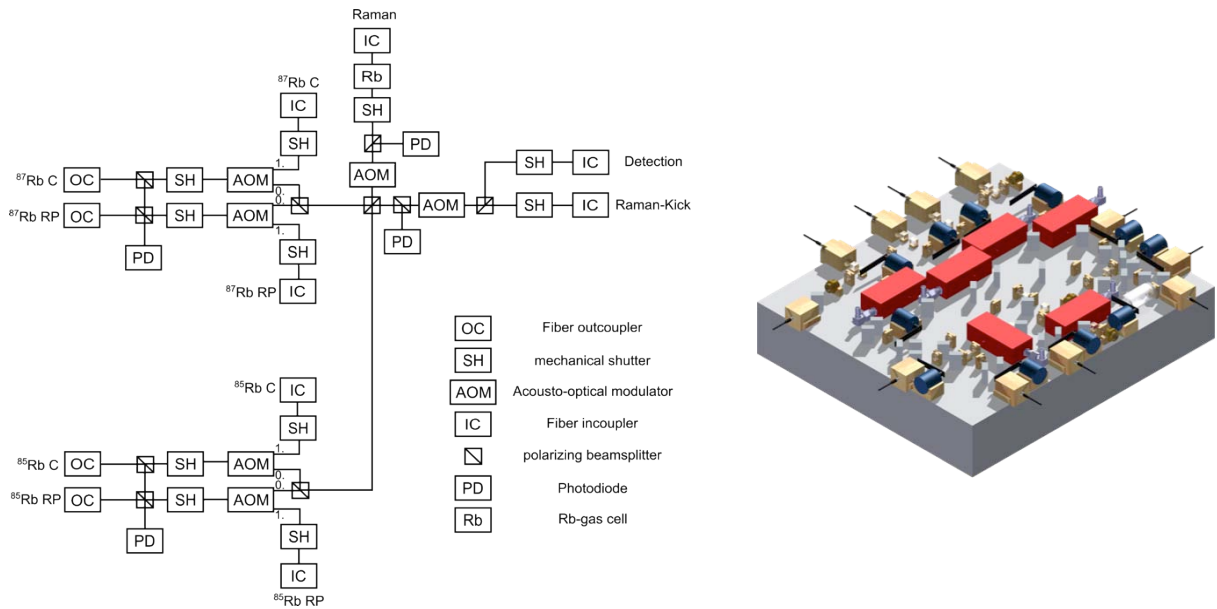

**Figure 21:** Schematic and CAD drawing of the switching module subsystem. It is based on Zerodur optical bench technology and comprises passive and active components for laser beam manipulation and distribution. Together with a fiber-based distribution module (not shown here), it forms the AI-SDM. (c) University of Hamburg.

The AI-SDM has overall dimensions of 400 x 370 x 125 mm$^3$, a total mass of 14.02 kg (16.82 kg incl. 20% component level margin), a total average power consumption of 6.02 W (7.22 W incl. 20% component level margin) and a peak power of 17.20 W (20.64 W incl. 20% component level margin).

## 6.4 Laser System Technology Readiness Estimation

The laser system includes a telecom technology based reference and optical dipole trap laser (AI-ROL), micro-integrated diode laser modules at a wavelength of 780 nm (AI-DLP) and a switching and distribution unit (AI-SDM) mainly based on free-space optics using a specific assembly-integration technology. The AI-ROL is mainly based on PHARAO heritage, yielding to TRL 5 of most of the components except the fiber amplifer (EDFA) with TRL 4. The AI-DLP and the AI-SDM modules as well as the spectroscopy unit within AI-ROL are based on QUANTUS/MAIUS heritage with a TRL between 4 and 5. Their design is optimized with respect to compactness, power consumption and mass and sounding rocket specific vibration tests were carried out.

## 7. Electronics Design and Software Architecture

This chapter describes a reference architecture for the electronics (Chapter 7.1) and the corresponding software (Chapter 7.2) needed for AI operation.

## 7.1 Electronics Units

The electronics consists of the following five functional units:

1. DMU (data management unit): This main digital control unit is based on the LEON processor developed under an ESA program – the LEON 2 processor (as e.g. used for all ESA instruments on the EarthCARE satellite) is available as a standalone device. Speeds up to 85 MHz are possible with the current version of the LEON device. The DMU also contains data acquisition electronics and housekeeping systems. It includes e.g. interfaces to photodiodes monitoring the fiber harness, to thermistors and to the CCDs. This unit controls all the other electronics units and the overall payload.

2. Magnetics Drive: This is a low voltage power supply providing low noise current drives for the magnetic field generation. These need to balance the need for low noise but high efficiency. A mixture of switched mode and linear control is required to minimize power consumption. The DMU instructs the magnetic drive to set a coil drive on/off at a specified current.

3. Low noise RF generator: It provides a 100 MHz signal as a reference for other AI electronics subsystems and up converted to 7 GHz and 3 GHz as input for the beam splitter phase locks and the micro wave antenna inside the physics package. DDSs (direct digital synthesis) allow tuning of the micro wave signals for parameter optimizations. Stable references and microwave chains [56] typically used in atom interferometry experiments are sufficient.

4. Laser Control: The laser system requires low noise current supply with a high bandwidth control loop, thermal control loops to maintain the stability of the lasers and frequency control loops for reference laser locking and offset frequency locking. Because of the loop bandwidth and low noise requirements the laser control electronics is mainly realized in the analog domain. Also data for housekeeping and closing control loops are measured (laser temperature, power amplifier temperature, supply current, frequency lock).

5. Ion Getter supply: The Ion getter pump requires a high voltage supply, in the order of 5 kV. This is powered independently from the other electronics functions as it requires to be powered during launch. The DMU only monitors its health.

Each of these units requires a level of redundancy, which is achieved internally. Figure 22 shows an overview of the system electronics. The technological maturity of each of the constituent parts is high – often with flight heritage, and almost all with mature design history. Because no model of the unit on system level has been built, the formal classification of the TRL is low (3 to 4). This, however, overestimates the program risk as a new technology development is at TRL4.

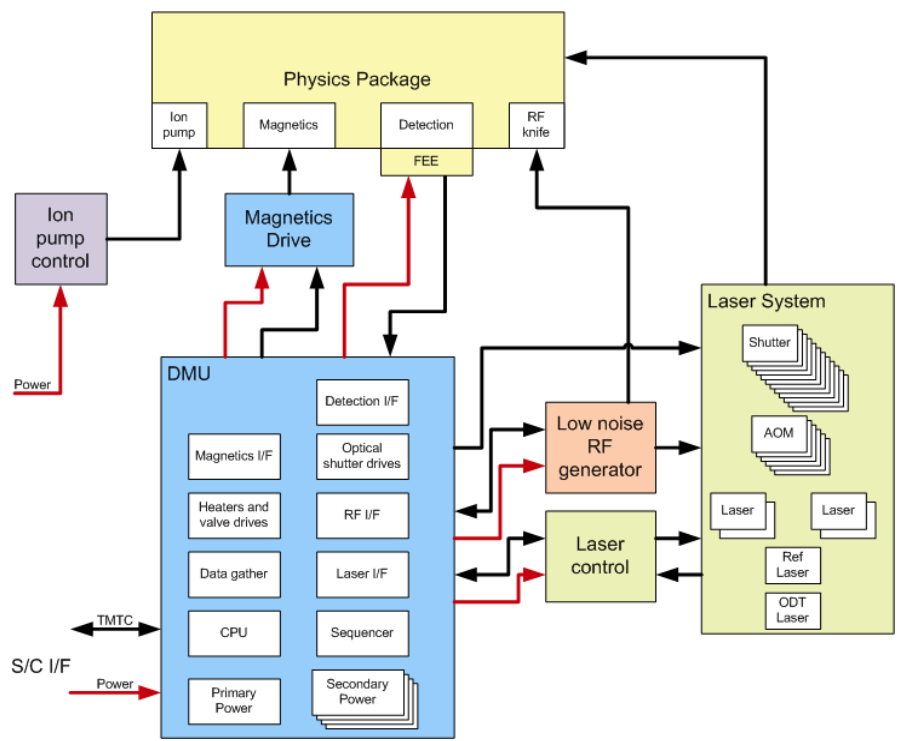

**Figure 22:** Electronics system overview.

## 7.2 Software Architecture

Developing software for the DMU implies using embedded techniques. The specific system requirements, not only software but electronic requirements, tie the software architecture to the hardware necessities. The system works in a time-constrained environment and performance is one of the most important goals.

### 7.2.1 Architecture and design

The AI software for Command, Control and Data Processing consists of two separated subsystems: the so-called OnBoard Software, being the Experiment Operations and Control Software, running on the S/C Computer, also called OnBoard Computer (OBC), and the ICU Software running on the DMU.

The ICU Software is split into the Boot (or Basic) Software (BSW), in charge of the initialisation and troubleshooting of the DMU and the Application Software (ASW), based on RTEMS [54] operating system, implementing the required science computations, control and data management for the rest of AI electronics units.

*BSW* is stored in the PROM, providing the minimum functionality necessary to:

- assess and report on the overall DMU hardware health status,
- establish a reliable communication link with the OBC, implementing an adequate subset of the Packet Utilization Standard (PUS) protocols ( ECSS-E-70-41A),
- check and provide access to RAM and EEPROM memory (where the ASW shall be stored),
- allow remote patching of Application Software.

*ASW* is an extension of the BSW. ASW's functionality can be summarized in three main tasks:

1. Handling of the AI subsystems
    - reroute TC from OBC to AI subsystems,
    - reroute TLM (housekeeping) from AI subsystems to OBC,
    - power management,
    - data acquisition rerouted to OBC for scientific purposes.

2. Computation of science data
    - for controlling experiment sequences, like parameter optimization and sensor pictures processing,
    - for controlling AI subsystems,
    - for sending to OBC.

3. System monitoring, including health status and Onboard Monitoring Function, the standard service specified by CCSDS.

The output context diagram for the ASW is shown in Figure 23.

The main behavior required for AI software, from experimental and science point of view, besides of standard housekeeping and monitoring, can be grouped in 3 main blocks:

- *Experiment Management:* Atom interferometry experiments require the simultaneous action on several devices. The timings involved in the experiment sequences are considered a hard and critical requirement; this means that the whole software (running under a 33 MHz or 85 MHz CPU) must be able to manage critical time sequences (steps around μs and changes about nanoseconds). In this first design, despite using a real-time operating system, this is assumed to be unavailable in strict terms of software. The best approach is to implement the experiment sequence using dedicated hardware electronics (FPGA) where the management and control of the parameters definitions is left to software.

- *Parameter Optimization:* The parameters in the experiment sequences must be very precise in order to produce best results. These parameters need to be computed using some function parameter optimization technique. As the processing power of the DMU is moderate, a hybrid approach is foreseen: a preliminary optimization must be performed on ground, leaving to flight software to only further improve them.

- *Image processing:* CCD sensors produce images that must be managed by the AI software in order to be able to send them to ground when possible (depending on mass storage present in DMU), but also an on-board processing is requested to be implemented. Some filters and fit algorithms are needed in order to extract useful information. The preliminary approach is to use a dedicated FPGA to implement the filters' algorithms, then software will use it as a co-CPU.

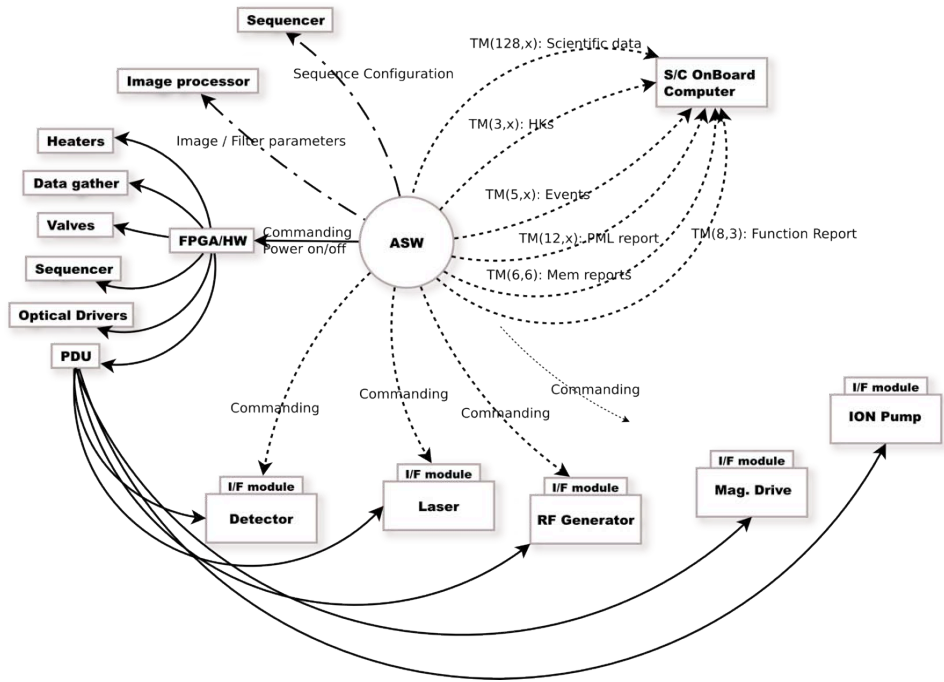

**Figure 23:** AI software output context diagram (Application Software).

### 7.2.2 Conventions, procedures, standards and quality

Regarding design methodology we have chosen the Ward-Mellor method [55], based on the well-known Yourdon structured analysis. It provides extensions taking into account real-time needs. It completely covers our needs for the whole DMU SW development. More heavyweight methodologies, like object-oriented developments using UML or RUP, are not justified for real-time applications as architecturally simple (or low-level) as ours.

Documentation and code is traced and versioned using appropriate tools for control versioning, issue tracking and requirement management as it is mandatory for computer engineering.

The main programming language to be used for developing the DMU SW is C, for BSW several parts are directly developed using the Leon CPU assembly language. These languages are best suited due to the need of access for underlying hardware at low level, and in order to ensure that the size and CPU consumption of the resulting executables is well within budget.

Quality assurance is an integral process enclosing all stages of the DMU software development life cycle. It relies on the ESA Software Engineering and Software Product Assurance and standards (ECSS–E–40 Part 2B), tailoring that standard at first stage of the software design phase.

Software testing and validation also follows the tailoring regarding quality. All the software produced is unit-tested by the same developer team and validated by external institutions/companies. Official test and validation software campaigns are planned prior to reach Qualification Review meeting and Acceptance Review.

## 8. Conclusion

In this paper, we presented a design of a high sensitivity dual species $^{85}$Rb/$^{87}$Rb atom interferometer for operations in space, as it was developed in the context of the STE-QUEST space mission dedicated to perform a quantum-test of the WEP with unprecedented precision. While the design was developed in order to meet the specific STE-QUEST requirements, the underlying technologies for realizing space atom interferometers are more general and can be applied and/or adapted to further atom interferometry applications in space. Recently, a $^{87}$Rb gravity gradiometer with a simultaneous production of BECs and interferometry for an interleaved operation mode was proposed [16]. The presented design is based on already well-proven technology used in drop-tower and parabolic flight experiments and optimized with respect to dimensions, mass and power consumption for space flight. Where possible, space-proven technology is foreseen. Current efforts aim at improving the technology readiness of the needed technology on component, subsystem and instrument level as most components and subsystems have been verified in a laboratory environment and basic functional performance as well as critical functions have been demonstrated.

In a first step, a (drop-tower operated) demonstrator of the detailed atom interferometer design will be set up, demonstrating (i) the simultaneous production of a $^{87}$Rb and $^{85}$Rb BEC with $10^6$ atoms each in a microgravity environment in less than 10 s, (ii) the simultaneous preparation of samples with a temperature equivalent of 70 pK by using shallow magnetic/ODT potentials and delta kick cooling, and (iii) the common-mode noise suppression of a dual species atom interferometer with effective wave vector matching. Furthermore, the demonstrator also aims at verifying the capability of the detection system to perform shot noise limited atom number detection and the calibration scheme used for control of the Center of Mass position. By increasing the technology readiness, the demonstrator will pave the way for exciting physics possible with atom interferometers in space.

*Acknowledgements*

This work was supported by the German space agency Deutsches Zentrum für Luft- und Raumfahrt (DLR) with funds provided by the Federal Ministry of Economics and Technology under grant numbers 50 OY 1302, 50 OY 1303, and 50 OY 1304, the German Research Foundation (DFG) by funding the Cluster of Excellence "Centre for Quantum Engineering and Space-Time Research (QUEST)", the French Space Agency Centre National d'Etudes Spatiales, and the European Space Agency (ESA). Lluis Gesa, Ignacio Mateos and Carlos F. Sopuerta acknowledge support from Grants AYA-2010-15709 (MICINN) and 2009-SGR-935 (AGAUR). Kai Bongs acknowledges support from UKSA for the UK contribution. Baptiste Battelier, Andrea Bertoldi and Philippe Bouyer thank the "Agence Nationale pour la Recherche" for support within the MINIATOM project (ANR-09-BLAN-0026). Wolf von Klitzing acknowledges support from the Future and Emerging Technologies (FET) programme of the EU (MatterWave, FP7-ICT-601180).